\documentclass[conference,compsoc]{IEEEtran}
\IEEEoverridecommandlockouts

\ifCLASSOPTIONcompsoc
  \usepackage[nocompress]{cite}
\else
  \usepackage{cite}
\fi

\ifCLASSINFOpdf
\else
\fi

\usepackage{cite}
\usepackage{amsmath,amssymb,amsfonts}
\usepackage{graphicx}
\usepackage{textcomp}
\usepackage{xcolor}
\def\BibTeX{{\rm B\kern-.05em{\sc i\kern-.025em b}\kern-.08em
    T\kern-.1667em\lower.7ex\hbox{E}\kern-.125emX}}

\usepackage[all]{nowidow}
\usepackage[linesnumbered,lined,vlined,ruled,commentsnumbered,noend]{algorithm2e}
\colorlet{RED}{red}
\usepackage{listings}
\usepackage{url}

\usepackage{xspace}
\usepackage{multirow}
\usepackage{subfloat}
\usepackage{balance}
\usepackage{ctable}
\usepackage{listings}
\usepackage{color}
\usepackage[nobiblatex]{xurl}

\usepackage{multirow}

\usepackage{algpseudocode}
\usepackage{caption}

\usepackage{cleveref}
\usepackage{acronym}
\usepackage{fancybox}
\usepackage{verbatim}
\usepackage[normalem]{ulem}
\usepackage{float}
\usepackage{newfloat}
\usepackage{mdframed}

\usepackage{enumitem}
\usepackage{booktabs}
\usepackage{pifont}

\usepackage{subfigure}
\usepackage{ulem}
\usepackage[flushleft]{threeparttable}
\usepackage{booktabs}
\usepackage{tabularray}
\usepackage{ifsym}

\definecolor{red}{RGB}{255,0,0}
\definecolor{green}{RGB}{18,220,168}

\newcommand{\eg}{e.g.,\xspace}

\usepackage{pifont}

\newcommand{\del}[1]{\textcolor{red}{}}

\newcommand{\eat}[1]{}

\acrodef{mlaas}[MLaaS]{Machine Learning as a Service}
\acrodef{sdm}[SDMs]{Stateful Defense Models}
\acrodef{qpa}[QPA]{Query Provenance Analysis}
\acrodef{dnn}[DNNs]{Deep Neural Networks}
\acrodef{gnn}[GNNs]{Graph Neural Networks}
\acrodef{sota}[SOTA]{state-of-the-art}
\acrodef{oars}[OARS]{Oracle-guided Adaptive Rejection Sampling}
\acrodef{asr}[ASR]{Attack Success Rate}
\acrodef{pas}[PAS]{Provenance Anomaly Score}
\acrodef{fpr}[FPR]{False Positive Rate}
\acrodef{ttd}[TTD]{Time-To-Detect}
\acrodef{ml}[ML]{Machine Learning}
\acrodef{rnn}[RNN]{Recurrent Neural Network}
\acrodef{lstm}[LSTM]{Long Short-Term Memory}
\acrodef{lbp}[LBP]{Local Binary Pattern}

\newlength{\MaxSizeOfLineNumbers}%
\definecolor{keywordcolor}{rgb}{0.8,0.1,0.5}
\definecolor{lightlightgray}{gray}{.96}
\definecolor{lightgray}{gray}{.925}
\definecolor{medlightgray}{gray}{0.7}
\definecolor{medgray}{gray}{0.4}
\definecolor{darkgray}{gray}{0.35}
\definecolor{nearblack}{gray}{0.15}

\crefname{component}{Component}{Components}

\newcommand{\distance}{8pt}
\begin{document}
\title{Query Provenance Analysis: Efficient and Robust Defense against \\Query-based Black-box Attacks}

\author{\IEEEauthorblockN{
Shaofei Li\IEEEauthorrefmark{1},
Ziqi Zhang\IEEEauthorrefmark{2}\IEEEauthorrefmark{4}\thanks{\IEEEauthorrefmark{4}Co-corresponding authors.}, 
Haomin Jia\IEEEauthorrefmark{3},
Yao Guo\IEEEauthorrefmark{1},
Xiangqun Chen\IEEEauthorrefmark{1},
Ding Li\IEEEauthorrefmark{1}\IEEEauthorrefmark{4}
}
\IEEEauthorblockA{
    \IEEEauthorrefmark{1}Key Laboratory of High-Confidence Software Technologies (MOE), School of Computer Science, Peking University\\
\IEEEauthorrefmark{2}University of Illinois Urbana-Champaign\\
\IEEEauthorrefmark{3}School of Electronics Engineering and Computer Science, Peking University\\
\{lishaofei,yaoguo,cherry,ding\_li\}@pku.edu.cn,\\
ziqi24@illinois.edu, stevenjia@stu.pku.edu.cn
}
}

\maketitle

\begin{abstract}
Query-based black-box attacks have emerged as a significant threat to machine learning systems, where adversaries can manipulate the input queries to generate adversarial examples that can cause misclassification of the system. To counter these attacks, researchers have proposed \ac{sdm} such as  BlackLight and PIHA, which can reject queries that are ``similar'' to historical queries. However, recent studies show that existing approaches are vulnerable to a stronger adaptive attack, \ac{oars}. \ac{oars} can be easily integrated with existing attack algorithms to evade the \ac{sdm} by generating queries with fine-tuned direction and step size of perturbations utilizing the leaked decision boundary from the \ac{sdm}. 

In this paper, we propose a novel approach, \ac{qpa}, for defending against query-based black-box attacks robustly (against both non-adaptive and adaptive attacks) and efficiently (in real-time). 
Our key insight is that, instead of focusing on individual queries, utilizing features from the query sequence (termed \textit{query provenance}) can distinguish malicious queries from benign queries more effectively.
We construct a query provenance graph to capture the relationship between a new query and prior historical queries, and then design efficient algorithms to detect malicious queries based on the query provenance graphs. We evaluate QPA on four datasets against six query-based attacks and compare QPA with state-of-the-art SDM defenses. The results show that \ac{qpa} outperforms the baselines regarding defense robustness and efficiency on both non-adaptive and adaptive attacks. 
Specifically, \ac{qpa} reduces the \ac{asr} of \ac{oars} to 4.08\%, which is roughly 20$\times$ lower than the baselines. Moreover, \ac{qpa} achieves higher throughput (up to 7.67$\times$) and lower latency (up to 11.09$\times$) than baselines.

\end{abstract}

\IEEEpeerreviewmaketitle

\section{Introduction}

\ac{ml} is extensively employed in various critical applications~\cite{wang2021deep,6248074, liu2020computing}.
However, recent research has revealed that \ac{dnn} are vulnerable to adversarial examples~\cite{ilyas2018black,brendel2018decisionbased,papernot2017practical,szegedy2013intriguing}, which are maliciously crafted inputs with small perturbations and can cause the \ac{dnn} to behave abnormally. 
Existing attack strategies can be categorized into two types: white-box attacks~\cite{8835245,tashiro2020diversity} and black-box attacks~\cite{7958568,8429311}. Compared to white-box attacks, which require full knowledge of the victim model, black-box attacks only require query access and output confidences or predicted labels. Thus, black-box attacks are more practical in real-world scenarios, such as \ac{mlaas}~\cite{amazonrekognition, clarifai, platerecognizer}. In \ac{mlaas}, the model users can only query the \ac{dnn} for outputs. 
Researchers have proposed many query-based black-box attacks to construct adversarial examples~\cite{ilyas2018black, andriushchenko2020square, brendel2018decisionbased,chen2020hopskipjumpattack, li2020qeba, maho2021surfree,bounceattack}. These attacks can be very effective and achieve nearly 100\% success rate~\cite{yu2020cloudleak, juuti2019prada}. Thus, it is very important to protect cloud models against query-based black-box attacks.

To defend against query-based black-box attacks, researchers have proposed Stateful Defense Models (\ac{sdm})~\cite{10.1145/3385003.3410925, 9760120, 281294, 10.1016/j.future.2023.04.005}. The insight of SDMs is that query-based attacks require querying the \ac{dnn} with multiple similar inputs with bounded perturbations. Thus, \ac{sdm} can detect adversarial examples by comparing the similarity of each new query instance with the previous query history, identifying the most similar queries (with a pre-defined threshold).
Although existing \ac{sdm} can detect attacks in certain scenarios effectively, they can be defeated with more sophisticated adaptive attacks such as the Oracle-guided Adaptive Rejection Sampling (OARS) attack ~\cite{10.1145/3576915.3623116}. Unlike previous adaptive attacks that apply simple transformations to input queries, OARS employs adaptive methods to learn the decision boundary of \ac{sdm} based on feedback. Thus OARS can fine-tune the attack algorithm during the attack process to make the attack queries less similar to the historical queries, in order to evade the detection of \ac{sdm}. Consequently, it can achieve a 99\% or higher \acf{asr}.

The main reason that existing \ac{sdm} are vulnerable to OARS is that \ac{sdm} primarily rely on the \textit{individual features of one incoming query} and the decision of \ac{sdm} leaks information about the decision boundary. This weakness allows the adversaries to craft attack samples against the decision boundary. Existing \ac{sdm} leverage the similarity of the closest historical query as an individual feature, while overlooking the similarity information of other historical queries. Thus, adaptive attacks can find the most similar queries that will not trigger the detection scheme of \ac{sdm}. Moreover, simply lowering the threshold of existing \ac{sdm} to enhance their robustness is also impractical, as it cannot prevent stronger adaptive attacks, while potentially increasing the false positive rate. Consequently, a more robust defense method is required to counteract stronger adaptive attacks.

Our key insight is that,  
\textit{instead of focusing only on individual queries, we can distinguish the relationship between queries more effectively by utilizing features from the query sequence and joint analysis of all historical queries.}
Adversary attack sequences can form unique patterns due to the practice of consecutive construction of malicious query sequences with small perturbations, while benign query sequences lack such patterns as they are typically unrelated to each other.
Although adaptive attacks, such as \ac{oars}, can evade single-point detection used by \ac{sdm}, they cannot conceal the unique features of the entire query sequence. As a result, \ac{oars} can evade existing \ac{sdm} by finding the most similar queries that will not trigger the \ac{sdm}, but it cannot hide the similarity among instances in the query sequence. Therefore, we can detect query-based black-box attacks based on the sequence features.

Inspired by traditional provenance-based work in the system security area, this paper proposes \acf{qpa}, which leverages the sequence feature of query instances, termed \textbf{query provenance}, to defend against query-based black-box attacks robustly and efficiently. We represent the query provenance as a graph structure, termed \textbf{query provenance graph}, which captures the relationship between a new query and prior historical queries. We then rely on the query provenance graphs to detect malicious queries generated by adversaries.

In order to construct query provenance graphs, we first integrate similarity and structural features to form query provenance, which is robust against noisy data.
For attack queries, the query provenance is the preceding query sequence involved in constructing the query instance. Normal (benign) queries do not have a relationship with other queries and, thus, do not have the unique provenance feature as malicious sequences. Based on this observation, we link the new query with the most similar prior queries when a new query arrives, to construct the query provenance graph. The weight is their similarity. 

Leveraging the provenance graph connected to each new query instance, we can then detect query-based attacks based on their unique patterns. Malicious queries tend to aggregate, exhibiting high sequence similarity and unique graph structures, while benign queries are typically scattered and lack such distinct patterns. Consequently, if the query provenance of a new query instance forms such a distinct pattern, it can be identified as a potential malicious query. This approach allows the system to adapt to chaotic environments, as the noisy benign data can be effectively filtered out by leveraging the feature of query provenance.

However, realizing \ac{qpa} is a non-trivial problem because the query provenance graph will introduce additional computational overhead, especially as the graph size increases with the arrival of new queries. Thus, the key challenge of \ac{qpa} is \textit{how to robustly (against both non-adaptive and adaptive attacks) and efficiently (in real-time) detect attack queries based on the query provenance graph}. 
To address this challenge, we propose two solutions. 
First, we design an anomaly detection algorithm based on a combination of statistics analysis and \ac{gnn} classifier. The statistics analysis can exclude most of the benign queries by similarity aggregation. Then, the \ac{gnn} classifier can filter out the false positives considering the provenance graph structure features.
Second, the attacks can be intermittent, meaning the attack procedure can last for a long time. \ac{qpa} needs to use the provenance of each query instance to detect the attack. However, it is impossible to keep all the query provenance information in memory. To mitigate this issue, we design a dynamic management strategy to maintain the suspicious provenance graphs and reset the storage of graphs periodically.

We then implement our detection tool as a \ac{qpa}-based SDM system, and evaluate \ac{qpa} on its defense capability on six non-adaptive attacks, as well as their adaptive versions with two adaptive strategies. The results demonstrate that our system can defend against query-based black-box attacks more robustly and efficiently than existing \ac{sdm}, such as BlackLight and PIHA. Our system achieves an extremely low \ac{fpr} of 0.04\% in chaotic environments. It also
reduces the \ac{asr} of \ac{oars} to 4.08\%, compared to 77.63\% and 87.72\% for BlackLight and PIHA.
Simultaneously, our system outperforms our baselines by achieving higher throughput (up to 7.67$\times$) and lower latency (up to 11.09$\times$).

We summarize our major contributions as follows:

\begin{itemize}[noitemsep, topsep=1pt, partopsep=1pt, listparindent=\parindent, leftmargin=*]
    \item We introduce the idea of utilizing the query sequence features (i.e., query provenance) as a robust and efficient feature to distinguish malicious queries from benign queries in query-based black-box attacks on DNNs.
    \item We propose query provenance analysis (QPA) to defend against black-box attacks through efficient detection of malicious queries based on the construction of query provenance graphs. 
    \item We evaluate \ac{qpa} on non-adaptive and adaptive attacks on four widely-used image datasets and demonstrate that it outperforms existing \ac{sdm} in terms of detection accuracy, robustness and efficiency.
    \item We release the source code of \ac{qpa} at \url{https://github.com/0xllssFF/QPA} to facilitate further research on adaptive attacks and \ac{sdm} in this area.
\end{itemize}

\section{Background and Motivation}

\subsection{Query-based Black-box Attacks}\label{sec:QBA} 

Query-based black-box
attacks are a class of adversarial attacks that only require query access to the
target model. The attackers can repeatedly query the target model with crafted
inputs and get the corresponding outputs. Considering the types of output
information, query-based black-box attack algorithms can be divided into two types: score-based
attacks and decision-based attacks. Score-based attacks~\cite{ilyas2018black, andriushchenko2020square} require the
attacker to know all the probabilities over the whole label space.
Decision-based attacks~\cite{brendel2018decisionbased, chen2020hopskipjumpattack, li2020qeba, maho2021surfree} only need the predicted labels. Both types of attacks start from an original input and iteratively perturb the input
to generate adversarial examples so that they can be falsely classified as either a
targeted or untargeted label. The average number of queries to a successful
attack is used to measure the efficiency of query-based black-box attack algorithms. A successful
attack usually requires hundreds to thousands of queries. The general pipeline
of query-based black-box attack algorithms consists of three steps: gradient estimation, step
execution, and boundary identification.

\noindent \ding{172}~\underline{Gradient Estimation:} 
Given an attack point $x_{t}$, each step starts from a Gaussian noise vector $\delta_{t}$ that has the same size as $x_{t}$. The attacker then queries the target model $f$ with $x_{t}+\delta_{t}$ to get the corresponding output. For each iteration of estimation, the attacker generates $n$ noise vectors and the gradient can be estimated by
\begin{equation}
    \nabla_{x}f(x_{t}+\delta_{t})\approx \frac{1}{n}\sum_{i = 1}^{n}L(f(x_{t} +
\delta_{t_{i}}), f(x_{t})),
\end{equation}
where $L$ is the loss function, while different algorithms have different
strategies.

\noindent \ding{173}~\underline{Step Execution:} 
Once the gradient has been estimated, the attack algorithms perturb the start point $x_{t}$ to $x_{t}^{\prime}$ by taking a step in the gradient's direction. The step size is typically initialized based on the current iteration round number and attack algorithms often employ the binary search algorithm to find an appropriate step size to improve convergence.

\noindent \ding{174}~\underline{Boundary Identification:} Some attack algorithms update the $x_{t}^{\prime}$ by linearly interpolating between itself and the original victim sample $x_{vic}$. This interpolation shifts $x_{t}^{\prime}$ closer to the decision boundary, thereby selecting a more favorable starting point for the subsequent iteration.
The interpolation procedure is conventionally guided by the binary search algorithm, which navigates the parameter space between $x_{t}^{\prime}$ and $x_{vic}$ to get closer to the decision boundary of the target model.

\subsection{Stateful Defense Models}\label{sec:bg_sdm} 
To defend against
query-based black-box attacks, researchers have proposed various \ac{sdm}~\cite{10.1145/3385003.3410925, 9760120, 281294, 10.1016/j.future.2023.04.005}. The key insight behind \ac{sdm} is that query-based black-box attacks typically generate many similar queries in gradient estimation and boundary identification stages. Thus,
we can detect query-based black-box attacks by detecting queries that exhibit high similarities.

The existing approaches are mostly \ac{sdm}
based on similarity between individual queries~\cite{10.1145/3385003.3410925,9760120,281294,10.1016/j.future.2023.04.005,10.1145/3576915.3623116}. These approaches compare ONE incoming query to each individual query in the history and mark the incoming query malicious if it is very similar (e.g., the similarity is higher than a predefined threshold) to another query in the history. However, the detection relying solely on individual similarity comparison has two significant limitations. 
First of all, it lacks stability, as it can introduce false positives because benign queries may show high similarity unexpectedly. 
Second, it is less robust, as relying on individual similarity is vulnerable to more advanced adaptive attacks.
Recent research has shown that advanced adaptive attacks can fine-tune the direction and step size of perturbations to ensure that the similarity between queries does not collide with the threshold-based decision boundary~\cite{10.1145/3576915.3623116}, which we will introduce in detail next.

\subsection{Adaptive Attacks}\label{sec:bg_adaptive}
Current individual similarity-based \ac{sdm} have been demonstrated to be susceptible to adaptive attacks. These attacks strategically manipulate the similarity between queries to avoid intersecting with the threshold-based decision boundary.
Generally, existing adaptive attacks can be divided into two categories: query-blinding strategy~\cite{10.1145/3385003.3410925} and adapt-and-resample strategy~\cite{10.1145/3576915.3623116}.

The query blinding strategy aims to ascertain the output of the target model while concealing the initial query from the defense model. This strategy employs common image transformation techniques such as rotation, translation and brightness adjustment so it can be easily applied without modifying the attack algorithm itself. However, this strategy often influences the attack's efficacy, potentially resulting in a successful evasion at the expense of the attack's failure~\cite{10.1145/3576915.3623116}. The adapt-and-resample strategy, \ac{oars}, is more sophisticated and does not merely transform images. It utilizes the leaked information of the defense model to perform rejection sampling that can perceive the discriminant boundary and adaptively adjust the attack parameters to escape detection. It utilizes binary search to find the optimal step size and noise distribution to generate queries that are most similar to the previous queries but will not cause collision. It is a general approach that can be applied to any attack consisting of the above three attack elements. The \ac{oars} is an advanced adaptive attack compared with  query blinding because it leverages the leaked information of the defense model to adjust the attack direction adaptively but the query blinding strategy is aimless.

Although different adaptive strategies have different attack performances, they all bypass the detection of existing \ac{sdm} by circumventing the threshold-based decision boundary. Therefore, the individual feature-based \ac{sdm} are not robust enough to detect adaptive attacks. To address this challenge, we propose to utilize sequence features instead of individual features to enhance the detection capability of \ac{sdm}. We will demonstrate why sequence features are more robust using an example in the following subsection.

\subsection{Key Insight}\label{sec:insight}

Based on the above limitations of existing \ac{sdm}, we propose a novel approach that leverages query sequence features to bolster the robustness of \ac{sdm}. Our approach is based on the observation that the attack query sequences can be aggregated in a highly structured pattern, while benign query sequences do not exhibit such patterns.

As detailed in Section~\ref{sec:QBA}, typical query-based black-box attack algorithms create adversarial examples by iteratively modifying inputs. The attacker first crafts multiple queries from a current attack point to estimate the gradient (\ding{172}). Then the attacker updates the attack point by taking a step in the direction indicated by the gradient to approach the boundary (\ding{173} and \ding{174}). 
Therefore, the attack sequence shows a spread (broad sampling of queries) and advance (progressive movement towards the boundary) pattern in each iteration. As the attack proceeds, the query sequence forms a highly organized tree structure with the spread-and-advance pattern.
This pattern is based on the similarity-based historical relationships among queries and is consistent across different attacks. 
While not all attack algorithms incorporate all three steps, and new attack algorithms may emerge, the attack sequence retains the aggregation property that forms a unique pattern.
Such sequence features remain in adaptive attacks like \ac{oars} because \ac{oars} also follows the three-element pipeline. The generated queries by \ac{oars} also exhibit higher similarity than benign queries. As a result, these sequence features can still distinguish attack patterns and enhance robustness against sophisticated adaptive strategies.

\begin{figure}[!t]
    \centering
    \includegraphics[width=0.49\textwidth]{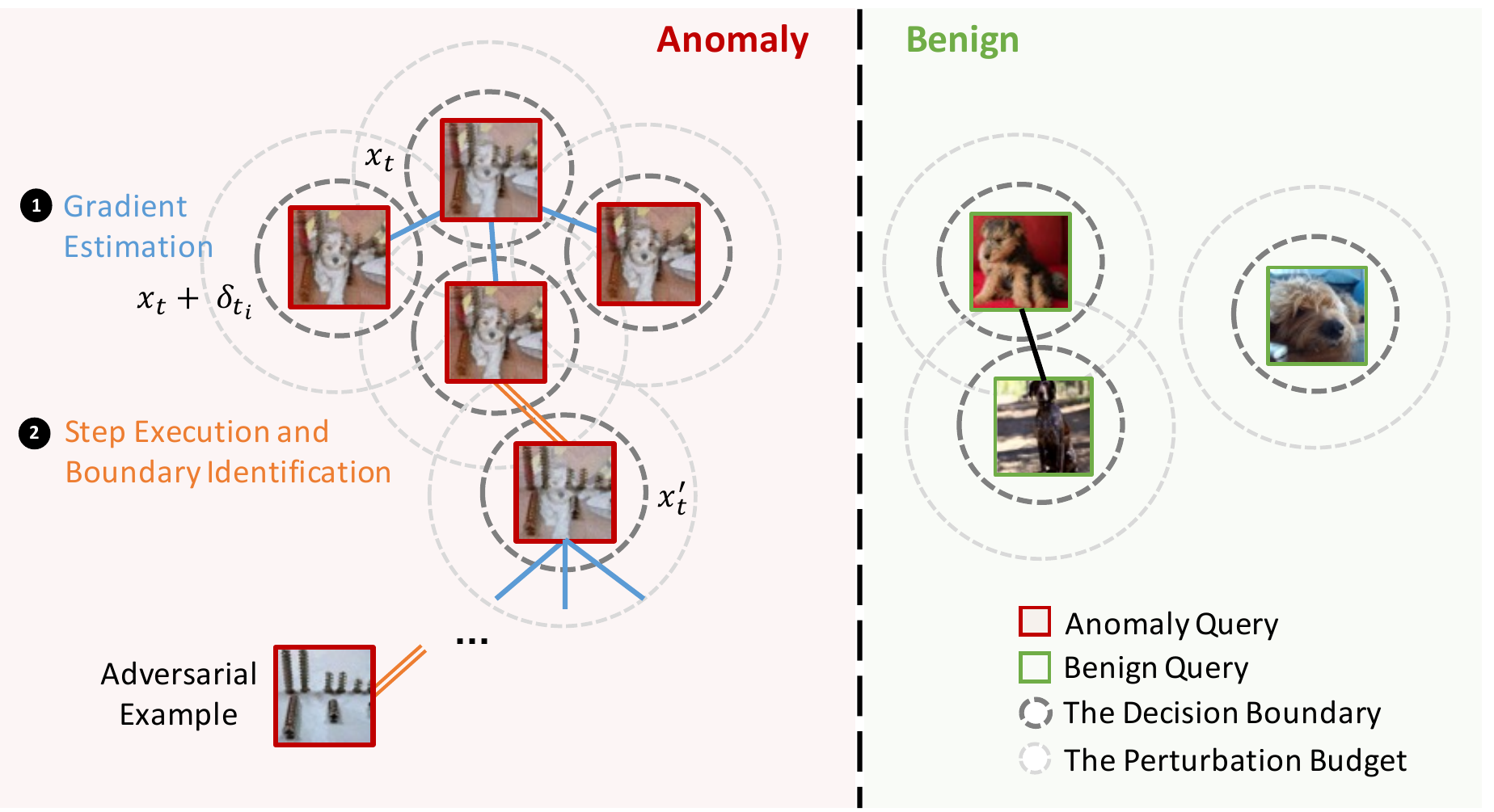}
    \caption{A motivating example of query provenance graph. $x_{t}$ indicates the start point, $x_{t} + \delta_{t}$ indicates the perturbed queries, and $x_{t}^{\prime}$ indicates the start point of the next iteration. The attack query sequence forms a highly organized graph structure while the benign query sequence exhibits random aggregation. 
    Existing \ac{sdm} employ a threshold-based decision boundary to detect the attack queries, which is vulnerable to \ac{oars} that can generate queries outside the decision boundary while inside the perturbation budget.
    }
    \label{fig:motivation}
\end{figure}

\section{Query Provenance Graph}\label{sec:bg_qpa}

Provenance analysis is widely used in system intrusion detection and investigation~\cite{king2003backtracking,HOLMES,nodlink,Han2020}. It leverages the relationship between system entities to detect and investigate the attack.
Recognizing that sequence features encapsulate the historical relationships among queries, we refer to this type of feature as \textbf{{query provenance}} and organize the history queries as \textbf{{query provenance graphs}}. 
A query provenance graph is an undirected graph $G = (V, E)$ where $V$ is the set of nodes representing the queries and $E$ is the set of edges with weights $sim(F(x_{i}), F(x_{j}))$ that measures the similarity between two queries $x_{i}$ and $x_{j}$, where $F(x)$ is the feature extraction function.
The graph is incrementally constructed as new queries arrive, by adding the new query as a node and linking it to the most similar query. 

Figure~\ref{fig:motivation} illustrates the query provenance graph with an adversarial example generated by the HSJA attack~\cite{chen2020hopskipjumpattack} with the \ac{oars} strategy. 
Each node represents a query and we organize the history queries as query provenance graphs by connecting the newly arrived query to the most similar previous queries. 
The left part of Figure~\ref{fig:motivation} represents a consecutive process to generate an adversarial example, which misleads the model to classify a screw as a dog. On the right, benign queries are randomly aggregated. 
For the attack query sequence, the graph progressively expands as the attack unfolds, culminating in successfully generating an adversarial example. Initially, the attacker crafts multiple queries $x_{t} + \delta_{t}$ from a current attack point $x_{t}$ to estimate the gradient. Therefore, the query sequence forms a subgraph with the origin point as the center. After estimating the gradient, the attacker updates the attack point by taking a step in the direction of the gradient. Then the algorithm modifies the newly generated point by interpolating a line between it and the original victim sample to approach the boundary and establish the starting point $x_{t}^{\prime}$ of the next iteration.
This step can be regarded as a forward move on the graph. These three steps are then repeated until the adversarial example is successfully generated.

The query provenance graph is robust against adaptive attacks (i.e., the OARS strategy). \ac{oars} can adjust the direction and step size of perturbations to generate queries outside the threshold-based decision boundary of existing \ac{sdm} and evade detection.
Using the query provenance graph, the attack query sequence delineates a highly organized tree structure, a feature that the \ac{oars} strategy can hardly circumvent. 
Conversely, the benign query sequence forms a random aggregation form, as shown on the right side of the figure.
Thus, the query provenance graph can capture the relationships between malicious queries and distinguish them from benign queries, which is effective even for chaotic data. 
We can then utilize the query provenance graph to design more robust \ac{sdm} to defend against adaptive query-based black-box attacks. The details of query provenance graph construction will be discussed in Section~\ref{sec:graphconstruction}.

\section{Threat Model and Design Goal}
\label{sec:threatmodel}

Our threat model follows prior defense against query-based black-box
attacks~\cite{281294}. We consider the following assumptions in our threat model:

\noindent\textbf{Attackers' Ability.} 
The target \ac{dnn} are deployed as a service. The attacker has no access to the training data and the architecture and parameters of the target model. 
The attacker can only query the target model with valid inputs and get the corresponding outputs, a full probability distribution across labels or only the predicted label. The attacker can only generate adversarial examples by querying the target model and cannot utilize other information.
The attacker can switch the attack between multiple user accounts and the cost is negligible.

\noindent\textbf{Attackers' Goal.}
The attacker aims to generate adversarial examples that can cause the target
model to misclassify the original inputs. The attacker can use any query-based
black-box attack algorithm to generate adversarial examples to achieve a high
success rate and low query cost.

\noindent\textbf{Defenders' Ability.} The defender has access to the inputs
and outputs of the target model. The defender does not need to know the model architecture or internal
states. The defender has limited resources for defense, such as computation
and storage, which means it is unacceptable if resource consumption continuously increases.

\noindent\textbf{Defenders' Goal.}
Our goal to defend against query-based black-box attacks is threefold:
\begin{itemize}[noitemsep, topsep=1pt, partopsep=1pt, listparindent=\parindent, leftmargin=*]
    \item \textbf{Robustness.} The defense model should be robust against both
    non-adaptive and adaptive attacks, especially for stronger adaptive attacks,
    \eg \ac{oars}.
    \item \textbf{Accuracy.} The defense model is designed to detect and defend against the generation of adversarial examples accurately, ensuring high coverage and a low false positive rate. It should be capable of identifying adversarial examples generated by various query-based black-box attacks. Furthermore, the model is designed to adapt to real-world scenarios, where benign queries significantly outnumber attack-related queries.
    \item \textbf{Efficiency.} The defense model should be efficient and
    introduce small latency in responding. The defense solution
    should be scalable to large-scale systems. It should also be able
    to handle a large number of queries in real time and introduce minimal
    latency to the response of queries.
    
\end{itemize}

\begin{figure*}[!t]
    \centering 
    \includegraphics[width=0.98\textwidth]{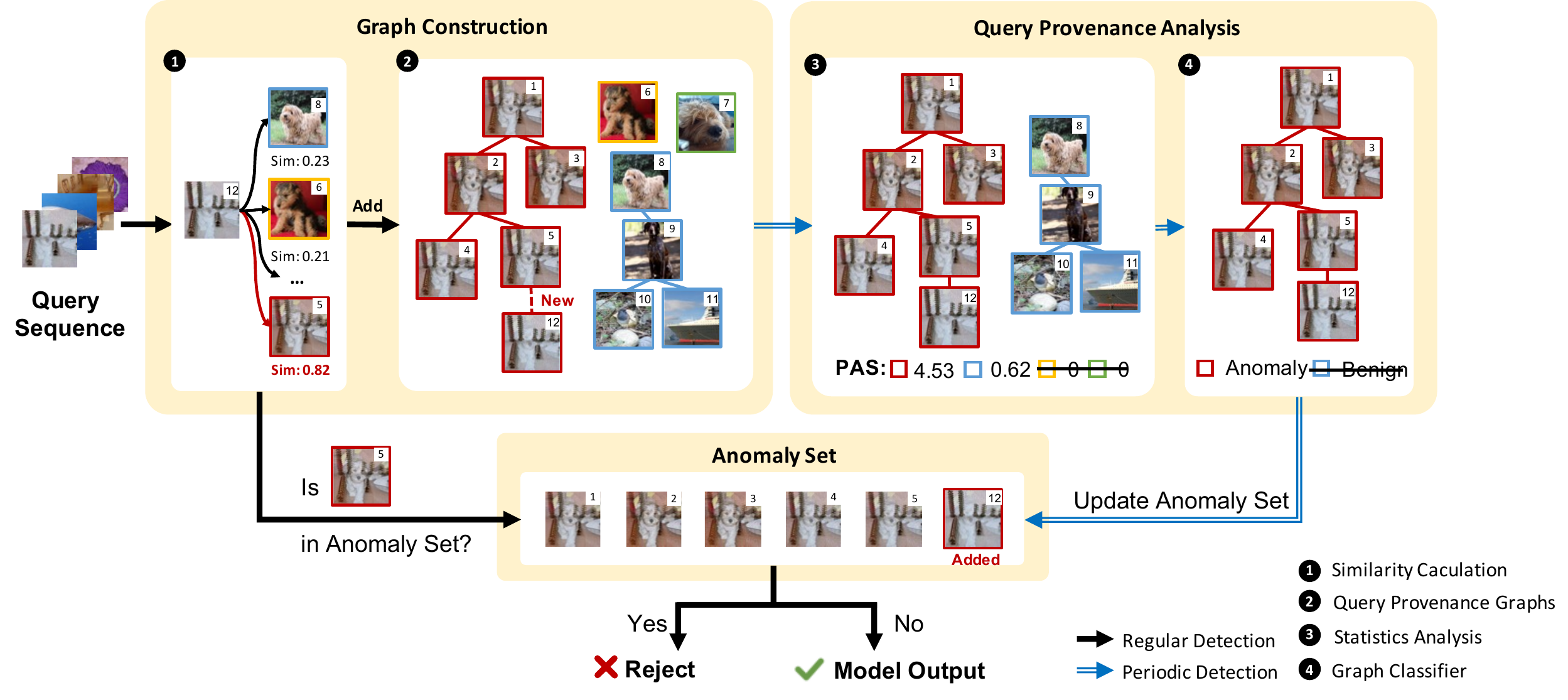}
    \caption{Workflow of our system with \ac{qpa}. The system receives the query stream as input and constructs the query provenance graph based on the similarity between queries. It analyzes the query provenance graph to update the anomaly set periodically. As a result, it rejects the malicious queries and returns model outputs only for benign ones. }
    \label{fig:overview}
\end{figure*}

\section{Design}\label{sec:design}

We propose to leverage query provenance for online query-based black-box attack detection, as depicted in Figure~\ref{fig:overview}. Our \ac{qpa}-based \ac{sdm} comprise two main components: graph construction and query provenance analysis. The former constructs the query provenance graph based on the similarity between queries, while the latter analyzes these query provenance graphs to detect malicious sequences. To detect long-running attacks in chaotic environments, we implement a two-phase detection mechanism: statistics analysis and graph classifier. The statistics analysis is lightweight and efficient, filtering out suspicious query provenance graphs. The graph classifier improves detection accuracy by classifying the outputs of the statistics analysis.
We also incorporate a dynamic management strategy of query provenance graphs, including graph eviction and reset. We adopt graph eviction to maintain a suspicious query history in memory and evict benign queries to the disk database. We also periodically reset graph storage to maintain system efficiency. The design of each component is detailed in the following subsections.

\subsection{Graph Construction}\label{sec:graphconstruction}

The graph construction component takes a query stream as input and organizes the query history into a graph structure. It builds the query provenance graph incrementally, utilizing the similarity between queries calculated from query features extracted by the feature extractor.  

\noindent \textbf{Similarity Calculation.} 
We utilize the feature extractor $F(x)$ from PIHA~\cite{10.1016/j.future.2023.04.005} as the default method to extract features from the query $x$ and compute the similarity between queries based on these features.
PIHA employs the Locality Sensitive Hashing (LSH) for feature extraction and preserves a fixed number of hash values as the query features.
It uses the Gaussian filter and \ac{lbp} algorithm as the feature extractor $F(x)$, as they are more resilient and efficient against adversarial perturbations compared to the BlackLight extractor. The similarity function is defined as $$sim(F(x_{i}), F(x_{j})) = \frac{|F(x_{i}) \cap F(x_{j})|}{|F(x)|},$$ where $|F(x_{i}) \cap F(x_{j})|$ denotes the number of common hash values between the features of $x_{i}$ and $x_{j}$ and $|F(x)|$ denotes the number of the hash values in the feature.

\noindent \textbf{Graph Initialization.} 
Attack queries tend to form highly structured query provenance graphs, a trait uncommon to benign queries.
This difference is crucial, allowing for effective classification of suspicious query provenance during the \ac{qpa} phase.
However, benign queries can introduce random aggregation into the query provenance graph, which may result in false positives. 
To mitigate this, we pre-initialize the graph with a small set of benign queries, typically 1,000, as distinct nodes.
This approach decreases the chances of benign queries linking within the same graph. These benign queries are chosen either from the benign set or those already processed by the target model. To maintain diversity among benign queries, we select them from different classes for certain datasets.

\noindent \textbf{Incremental Construction.}
The query provenance graph is constructed incrementally as the new query arrives.
A straightforward approach would be to add the incoming query as a node in the graph and connect it to the most similar queries.
However, this method may introduce false positives because the benign queries may have the highest similarity to attack queries with extremely small similarity,  complicating subsequent detection. 
To mitigate this issue, we set a threshold $T$ for the similarity based on the observation that the similarity between attack queries significantly exceeds that between entirely unrelated images. 
If the similarity between the incoming query and the most similar query falls below this threshold, we only add the incoming query as a node without connecting it with an edge. We set the threshold $T$ to the \textit{90th percentile} similarity between the benign queries used in the initialization. 
We compare the detection performance with and without $T$ in Section~\ref{sec:hyperparameters}.

\subsection{Query Provenance Analysis (QPA)} \label{sec:qpa}

The \ac{qpa} is used to analyze the query provenance graphs and detect malicious sequences. It receives the query provenance graphs as input and detects the anomalous graph as an anomaly set. \ac{qpa} detects anomalies based on the \ac{pas} and the structure of the query provenance graph. To guarantee the detection accuracy and efficiency, we adopt a two-phase detection mechanism: 
1) Statistics Analysis and 2) Graph Classifier. 
Statistics analysis calculates the \ac{pas} for each query provenance graph to filter out the suspicious graphs. 
Then the graph classifier utilizes \ac{gnn} to classify the outputs of the statistics analysis based on their \ac{pas} and graph structure. Instead of directly using the graph classifier, we use statistics analysis to filter out the suspicious graphs to reduce the computation overhead of the graph classifier.

\noindent\textbf{Statistics Analysis.}
In the statistics analysis, we calculate the \ac{pas} for each connected query provenance graph. The \ac{pas} is used to represent the sequence similarity of the query provenance graph. 
Considering that the attack query sequence is generated by continuously perturbing the input, the sequence similarity of the attack sequence is higher than that of the benign ones. Thus we can utilize the \ac{pas} to detect anomalies.

We calculate the \ac{pas} based on the similarity weight assigned to the edges between queries in the graph. This weight corresponds to the similarity between queries and their closest neighbors, established during the construction of the query provenance graph. The initial consideration might be defining \ac{pas} as the average of all the edges in the graph. However, this approach fails to robustly differentiate malicious queries from benign query sequences, especially for adapt-and-resample adaptive attacks. These attacks can identify directions and step sizes that minimize similarity, thereby reducing the distinctiveness of the similarity between queries. Thus the average similarity overlooks the scale feature of the query provenance graphs, which means that the nodes of the attack query provenance graph significantly outnumber those of benign ones. 
To distinguish between attack and benign queries more effectively, we define \ac{pas} as the sum of all edge weights in a connected graph, formally expressed as $PAS(G) = \sum_{e \in G}e_{weight}$. Even for adaptive attacks, the attack sequence will exhibit an significantly higher PAS compared to benign ones due to the high degree of aggregation.

To identify anomalies based on the \ac{pas}, we employ the Grubbs's test~\cite{Grubbs-test}, a statistical algorithm designed to detect outliers within a sample set. Specifically, the Grubbs's test can determine whether the largest value in a set of samples is an outlier.
Grubbs's test is defined for the following hypotheses:

$$G_{test} = \frac{Y_{max} - \overline{Y}}{s}$$ $$G_{critical} = \frac{N-1}{\sqrt{N}}\sqrt{\frac{t^{2}_{\alpha/N, N-2}}{N - 2 + t^{2}_{\alpha/N, N-2}}}$$
with $\overline{Y}$ and $s$ denoting the mean and standard deviation of the samples, $t_{\alpha/N,N-2}$ denoting the upper critical value of the t-distribution with $N - 2$ degrees of freedom and a significance level of $\alpha/N$. The null hypothesis, which posits the absence of outliers, is rejected at the significance level $\alpha$ if $G_{test} < G_{critical}$. In our system, we set the significance level $\alpha = 0.01$ for fewer false positives.

We selected Grubbs's test due to its lightweight nature and robustness in this scenario. We apply Grubbs's test to the query history iteratively until no outliers are identified. However, the detected outliers may contain false positives, which reduces detection accuracy. To address this, we incorporate a graph classifier to further categorize the outputs of the statistics analysis based on their \ac{pas} and graph structure.

\noindent\textbf{Graph Classifier.}
The graph classifier is designed to reduce the false positives of the statistics analysis, leveraging both \ac{pas} and graph structure. 
In real-world scenarios, benign queries significantly outnumber malicious queries, potentially leading to random aggregation in the query provenance graph. However, these aggregations do not form a highly structured graph like those formed by attack sequences. Based on this observation, we can combine \ac{pas} and graph structure to improve detection accuracy. 

We employ \ac{gnn} to learn graph embeddings and train a binary classifier for the classification. \ac{gnn} are powerful tools for capturing graph structure and node features. A straightforward approach is to use the original query provenance graph as the \ac{gnn}' input. 
However, since this graph contains only edge features, traditional \ac{gnn} such as Graph Convolutional Networks (GCN)~\cite{zhang2019graph} and Graph Attention Networks (GAT)~\cite{velickovic2017graph}, which rely on node features, may lead to suboptimal learning performance for embeddings.
To overcome this issue, we transform the original graph into its line graph~\cite{LineGraph}. A line graph represents the edges of the original graph as nodes, and the edges between these nodes represent the adjacency of the original graph. This transformation allows the line graph to capture both the graph structure and edge features simultaneously. We then feed the line graph into the classic GCN model to learn the graph embedding and perform classification. The loss function is defined as the cross-entropy loss between the predicted label and the ground truth.
The graph classifier model comprises three convolutional layers and one linear layer, offering a lightweight structure that introduces minimal latency for online detection.

The graph classifier is pre-trained offline and deployed for online detection. We impose a limit, denoted as $s$, on the number of nodes in the input graph. This constraint ensures that the structural characteristics of the graph are adequately represented, as these characteristics only become apparent when the graph reaches a certain scale. The parameter $s$ influences both the \ac{ttd}, which is defined as the number of queries to detection, and detection accuracy, which we will discuss in Section~\ref{sec:evaluation}.
The graph classifier is trained on randomly selected attack sequences and benign queries, with each dataset trained independently.
For anomaly samples, we randomly select five attack sequences for each attack algorithm and construct the query provenance graphs. We then save the graph as a sample for every $s$ queries. For benign samples, we randomly select the benign queries from the benign set. Given that benign queries exhibit less aggregation than attack queries, we save a graph sample for every 500 queries. 
Subsequently, we partition these data into training and testing sets at a ratio of 7:3. The graph classifier is trained using the training set, and the best-performing model on the testing set is saved for online deployment.

\subsection{Dynamic Management}
System efficiency is crucial for \ac{ml} services. The query provenance graph, which may expand indefinitely with incoming queries, can introduce significant computational overhead for detection. 
To counteract long-running adversarial attacks while ensuring the efficiency of \ac{ml} service, we design an eviction strategy to maintain the suspicious query history in memory while evicting the benign queries to a database on disk. This approach differs from previous \ac{sdm}, which retain all query features in memory. Instead, our strategy selectively retains only the more suspicious query provenance graphs, while efficiently offloading the benign ones to the database. We also design a graph reset strategy to reset the query provenance graph storage periodically. 
We integrate these dynamic management strategies in the periodic detection to maintain the integrity and performance of the system. 

\noindent \textbf{Graph Eviction.}
In our design of \ac{qpa}, the query provenance is organized as an undirected graph, with each connected query provenance graph assigned a \ac{pas} to represent the sequence similarity. A higher \ac{pas} indicates greater sequence similarity of the query provenance graph and the attack sequences will have significantly higher \ac{pas} than the benign ones.
Intuitively, we can preserve the graphs with the top $K$ highest \ac{pas} in the memory and evict the others to the database. This approach would reduce computational overhead for detection, as we only need to calculate the similarity with the suspicious query history. In real-world scenarios, malicious queries are far fewer than benign queries and exhibit aggregation in the query provenance graph, suggesting that the number of attack-related graphs is limited. Therefore, we can preserve only the top $K$ highest \ac{pas} graphs in the memory after the \ac{qpa} phase to facilitate subsequent detection. To ensure the detection capability, we preserve the graph structure in the database for integrated analysis between time windows by connecting the evicted nodes to the closest nodes in the database. The eviction procedure is asynchronous with online detection, ensuring it does not impact detection efficiency. We will discuss the impact of $K$ in Section~\ref{sec:hyperparameters}.

\begin{table}[t!]
    \scriptsize
    \centering
    \caption{Overview of the image classification task, including the dataset, the model architecture and accuracy.}
    \resizebox{0.4\textwidth}{!}{\begin{tblr}{
        cell{1-5}{1-3} = {c},
        hline{1,6} = {1pt}
    }
    \SetCell[r = 1, c = 1]{c} \textbf{Dataset}  & \SetCell[r = 1, c = 1]{c} \textbf{Model Architecture} & \SetCell[r = 1, c = 1]{c} \textbf{Model Accuracy} \\  \hline
    MNIST & 6 Conv + 3 Dense & 99.36\%\\ \hline
    CIFAR-10 &ResNet20 & 91.73\% \\ \hline
    ImageNet &ResNet152 & 78.31\%\\ \hline
    CelebaHQ &ResNet152 & 89.55\%\\ 
\end{tblr} }
    \label{tab:classification}
\end{table}

\begin{table*}[t!]
    \scriptsize
    \centering
    \caption{The configurations and performance of six non-adaptive attacks without any defense. ``ASR'' indicates the attack success rate and ``Avg \# of queries'' indicates the average number of queries needed to generate an adversarial example.}
    \resizebox{0.98\textwidth}{!}{\begin{tblr}{
        cell{1-2}{1-11} = {c},
        cell{2-8}{1-2} = {l},
        cell{2-8}{3} = {c},
        cell{2-8}{4-11} = {r},
        hline{1,9} = {1pt},
        stretch=0
    }
    \SetCell[r = 2, c = 1]{c} \textbf{Attack}  & \SetCell[r = 2, c = 1]{c} \textbf{Targeted} & \SetCell[r = 2, c = 1]{c} \textbf{Distance} & \SetCell[r = 1, c = 4]{c} \textbf{\ac{asr}} & & & & \SetCell[r = 1, c = 4]{c} \textbf{Avg \# of queries} & & &\\ \hline
    & & & MNIST&CIFAR10 &ImageNet & CelebaHQ & MNIST& CIFAR10 & ImageNet & CelebaHQ\\  \hline
    NES & Targeted &$L_{\infty}$ & 40\% & 100\% &99\% & 100\%& 49570 &707 & 14324 & 6737\\ \hline
    Boundary & Untargeted &$L_{2}$ & 73\% & 100\%& 100\%& 100\% & 15190 & 605 & 6468 & 730\\ \hline
    Square & Untargeted &$L_{2}$& 98\% & 100\%& 100\%& 100\% & 5731 & 390 & 346 & 632 \\ \hline
    HSJA & Targeted &$L_{2}$ & 82\% & 100\%& 100\% &100\% &1890 & 1246& 13237& 7956\\ \hline
    QEBA & Targeted &$L_{2}$ & 98\% & 99\%&  53\%&100\% & 2194& 2620 & 15425 & 3948\\ \hline
    SurFree & Untargeted &$L_{2}$ & 74\% &100\% &  100\% & 100\% & 10344&167 & 645 & 191\\ 
\end{tblr} }
    \label{tab:NonAdaptiveAttack}
\end{table*}

\noindent \textbf{Graph Reset.} Limited by the resources for defense, the query provenance graph cannot grow infinitely. Therefore we reset the storage of graphs every 24 hours to maintain the efficiency of the system. We reset the query provenance graphs by purging the memory and database, then reinitializing the query provenance graph with the benign queries. 
However, it may provide an opportunity for long-running attacks to escape detection. We will discuss the evasion of attacks in Section~\ref{sec:discussion}.

\section{Evaluation}\label{sec:evaluation}

We evaluate our system with six non-adaptive attacks and their adaptive versions with two adaptive strategies~\cite{10.1145/3576915.3623116, 10.1145/3385003.3410925} on four widely used datasets: MNIST~\cite{6296535}, CIFAR10~\cite{krizhevsky2009learning}, ImageNet~\cite{russakovsky2015imagenet} and CelebaHQ~\cite{karras2017progressive}. We compare our \ac{qpa}-based \ac{sdm} with two state-of-the-art \ac{sdm}: BlackLight~\cite{281294} and PIHA~\cite{10.1016/j.future.2023.04.005}. Our \ac{qpa}-based system is designed to achieve robustness and efficiency at the same time. Therefore, we proposed the following research questions to evaluate \ac{qpa}-based \ac{sdm}:

\begin{itemize}[noitemsep, topsep=1pt, partopsep=1pt, listparindent=\parindent, leftmargin=*]
    \item \textbf{RQ1:} Can \ac{qpa}-based \ac{sdm} detect and defend against query-based black-box attacks more effectively than existing \ac{sdm}?
    \item \textbf{RQ2:} Can \ac{qpa}-based \ac{sdm} detect query-based black-box attacks efficiently?
    \item \textbf{RQ3:} Do \ac{qpa}-based \ac{sdm} utilize query provenance to detect query-based black-box attacks?
    \item \textbf{RQ4:} How do the hyperparameters and each technique affect the detection performance of \ac{qpa}-based \ac{sdm}?
    \item \textbf{RQ5:} How do \ac{qpa}-based \ac{sdm} perform in real-world scenarios?
\end{itemize}

In this section, we will first introduce the experimental setup in Section~\ref{sec:experimentsetup}. Then, we will display the effectiveness of \ac{qpa} in front of both adaptive and non-adaptive attacks in Section~\ref{sec:effectiveness}. In Section~\ref{sec:efficiency} and Section~\ref{sec:insightverification}, we will display the efficiency of \ac{qpa} and verify our insight. In Section~\ref{sec:hyperparameters}, we will study the influence of hyperparameters and each component. 
At last, in Section~\ref{sec:realworld} we will evaluate the performance of \ac{qpa} in  real-world scenarios.

\subsection{Experimental Setup}\label{sec:experimentsetup}
Our experiments are focused on image classification tasks, which are the most widely used in previous work. To ensure evaluation fairness, we choose the datasets and targeted models that previous work has used~\cite{281294,10.1145/3576915.3623116}.
We separate the datasets into training, validation, and test sets at a ratio of 7:1:2. Images from the training set are selected as victims to generate attack query sequences for training \ac{qpa}. We select images from the validation set for tuning the hyperparameters and the test set to evaluate the defense performance.
We deploy our system and baselines on the server with 128-core Intel(R) Xeon(R) Platinum 8358 CPU and 488GB memory.

\noindent \textbf{Image Classification Tasks.}
We evaluate the defense performance of BlackLight and PIHA on the four most common datasets: MNIST, CIFAR10, ImageNet and CelebaHQ. For MNIST, we trained a LeNet-5~\cite{726791} convolutional network for digital number classification. For CIFAR10, ImageNet and CelebaHQ, we selected the targeted models that previous work has used~\cite{10.1145/3576915.3623116}. Table~\ref{tab:classification} lists the details of the datasets and targeted models.

\noindent \textbf{Baseline Configurations.}
As discussed in Section~\ref{sec:bg_sdm}, we choose two global query store design \ac{sdm} as our baselines: BlackLight~\cite{281294} and PIHA~\cite{10.1016/j.future.2023.04.005}. We use the open-source implementation and most of the default configurations of BlackLight and PIHA as described in their papers. For BlackLight, we configured its hyperparameters for different datasets as listed in the paper and set the same hyperparameters for CelebaHQ as ImageNet. We set the similarity threshold $q = 0.5$ for BlackLight.
For PIHA, we adjust the similarity threshold $q$, which is the proportion of identical hash values in a hash sequence, from 0.95 to 0.8 to achieve the best performance.

\begin{table*}[t!]
    \scriptsize
    \centering
    \caption{The detection performance of \ac{qpa} and two baselines on non-adaptive attacks. \textbf{BL} and \textbf{PIHA} indicate BlackLight and PIHA, respectively. The best results are in \textbf{bold}.}
    \resizebox{0.98\textwidth}{!}{\begin{tblr}{      
        cell{1-2}{1-9} = {c},
        cell{3-27}{1-2} = {l},
        cell{3-27}{3-14} = {r},
        hline{1,28} = {1pt}
        }
    \SetCell[r = 2, c = 1]{c} \textbf{Dataset}  & \SetCell[r = 2, c = 1]{c} \textbf{Attack} & \SetCell[r = 1, c = 3]{c} \textbf{Detection Coverage} & & & \SetCell[r = 1, c = 3]{c} \textbf{Detection Precision} & & & \SetCell[r = 1, c = 3]{c} \textbf{FPR} & & &  \SetCell[r = 1, c = 3]{c} \textbf{\ac{ttd}} & & \\  \cline{3-14}
    & & \textbf{BL} & \textbf{PIHA} & \textbf{\ac{qpa} } &  \textbf{BL} & \textbf{PIHA} & \textbf{\ac{qpa} } &  \textbf{BL} & \textbf{PIHA} & \textbf{\ac{qpa}} &  \textbf{BL} & \textbf{PIHA} & \textbf{\ac{qpa} }\\ \hline
    \SetCell[r = 6,c = 1]{r} \textbf{MNIST}  &  NES  
    & \textbf{0.98} & - & 0.95   
    & \textbf{1.00} & - & \textbf{1.00}
    & \textbf{0.00\%} & - & \textbf{0.00\%}
    & \textbf{2.1} & - & 20.3\\\hline 
    & Boundary 
    & 0.60 & - & \textbf{0.99}
    & 0.04 & - & \textbf{1.00}
    & 13.00\% & - & \textbf{0.00\%}
    & 2.4 & - & 20.3\\\hline 
    & Square  
    & 0.99 & - & \textbf{1.00}  
    & \textbf{1.00} & - & \textbf{1.00}
    & \textbf{0.00\%} & - & \textbf{0.00\%}
    & \textbf{2.0} & - & 18.5 \\ \hline
    & HSJA  
    & 0.94 & - & \textbf{0.97}  
    & \textbf{1.00} & - & \textbf{1.00}
    & \textbf{0.00\%} & - & \textbf{0.00\%}
    & \textbf{5.0} & - &  21.2\\\hline 
    & QEBA  
    & 0.83 & - &  \textbf{0.88} 
    & \textbf{1.00} & - & \textbf{1.00}
    & \textbf{0.00\%} & - & \textbf{0.00\%}
    & \textbf{8.1} & - & 23.4\\\hline 
    & SurFree 
    & 0.03 & - & \textbf{1.00}
    & 0.52 & - & \textbf{1.00}
    & 0.08\% & - & \textbf{0.00\%}
    & \textbf{6.0} & - & 20.1\\\hline 
    \SetCell[r = 6]{r}  \textbf{CIFAR10}   &  NES  
    & 0.97 & 0.98 & \textbf{0.99}  
    & \textbf{0.97} & 0.74 & 0.90
    & \textbf{0.04\%} & 0.36\% & 0.16\%
    & \textbf{2.0} & 2.5 & 18.9\\\hline 
    & Boundary 
    & 0.21  & 0.75 &  \textbf{0.94}
    & 0.86 & 0.61 & \textbf{0.96}
    & 0.05\% & 0.47\% & \textbf{0.03\%}
    &\textbf{2.5}  & \textbf{2.5} & 20.5\\\hline 
    & Square
    & 0.99 & 0.96 &  \textbf{1.00}
    & \textbf{0.97} & 0.85  & 0.93
    & \textbf{0.03\%} & 0.19\% & 0.10\%
    & \textbf{2.0} & \textbf{2.0} & 18.0\\\hline 
    & HSJA
    & 0.95 & \textbf{0.97} &  0.89 
    & \textbf{0.96} & 0.69 & 0.91
    & \textbf{0.04\%} & 0.44\% & 0.18\%
    & 4.9 & \textbf{2.3} & 19.3\\\hline 
    & QEBA 
    & 0.95 & \textbf{0.97} & 0.76  
    & \textbf{0.96} & 0.68  & 0.89
    & \textbf{0.04\%} & 0.46\% & 0.13\%
    & 5.1 & \textbf{2.5} & 19.3 \\\hline 
    & SurFree 
    & 0.07  & 0.53 &  \textbf{0.81}
    & 0.59  & 0.69 & \textbf{0.91}
    & 0.09\% & 0.22\% & \textbf{0.03\%}
    & 10.3 & \textbf{7.1} & 45.2 \\\hline 
    \SetCell[r = 6]{r}  \textbf{ImageNet}  & NES
    & 0.98 & 0.90 & \textbf{0.99}
    & 0.88 & 0.89 & \textbf{1.00}
    & 0.14\%     & 0.10\% & \textbf{0.00\%}
    & \textbf{2.1} & 9.3 & 17.7 \\\hline 
    & Boundary  
    & 0.03  & 0.35 &   \textbf{0.99}
    & 0.18  & 0.78 & \textbf{0.98}
    & 0.14\% & 0.10\% & \textbf{0.03\%}
    & 9.4 & \textbf{2.0} & 18.6\\\hline 
    & Square  
    & 0.98 & 0.98 &  \textbf{0.99}
    & 0.92 & 0.97 & \textbf{1.00}
    & 0.09\% & 0.04\% & \textbf{0.00}\%
    & \textbf{2.0} & \textbf{2.0} & 17.8\\\hline 
    & HSJA  
    & 0.95 & 0.95 & \textbf{0.98}
    & 0.87 & 0.91 & \textbf{0.98}
    & 0.14\% & 0.10\% & \textbf{0.00\%}
    & 6.2 & \textbf{5.7} & 17.7\\\hline 
    & QEBA  
    &0.94  &0.95  & \textbf{0.97}
    & 0.88 & 0.91 & \textbf{0.96}
    & 0.13\% & 0.09\% & \textbf{0.06\%}
    & 7.5 & \textbf{5.9} & 18.8\\\hline 
    & SurFree 
    & 0.05 & 0.06 & \textbf{0.99}  
    & 0.30  & 0.50 & \textbf{1.00}
    & 0.11\%  & 0.05\%  & \textbf{0.00}\%
    & 7.7 & \textbf{6.1} & 18.4\\ \hline
    \SetCell[r = 6]{r}  \textbf{CelebaHQ} & NES
    & 0.97 & 0.92  & \textbf{1.00}
    & 0.92 & 0.92 & \textbf{1.00}
    & 0.08\% & 0.08\% & \textbf{0.00\%}
    & 2.1 & 9.9 & 18.9 \\ \hline
    & Boundary
    & 0.01 & 0.69 & \textbf{0.99}
    & 0.09  & 0.90  & \textbf{1.00}
    & 0.09\% & 0.08\%  & \textbf{0.00\%}
    & 11.4  & \textbf{2.0} &  19.2\\ \hline
    & Square
    & 0.99  & 0.99  &\textbf{1.00}
    & 0.95 & 0.95 & \textbf{1.00}
    & 0.06\%  &  0.05\% &\textbf{0.00\%}
    & 2.0 & 2.2 & 18.7 \\ \hline
    & HSJA
    & 0.92 & 0.92 & \textbf{0.99}
    & 0.92 & 0.92 & \textbf{1.00}
    & 0.08\% & 0.07\% & \textbf{0.00\%}
    & 6.2 & \textbf{5.6} & 17.9 \\ \hline
    & QEBA
    & 0.92 & 0.85  & \textbf{1.00}
    & 0.92 & 0.91 & \textbf{1.00}
    & 0.08\% & 0.07\% & \textbf{0.00\%}
    & \textbf{6.1} & 11.8 & 18.0 \\ \hline
    & SurFree
    & 0.16 & 0.14 & \textbf{0.99}
    & 0.77  & 0.72 & \textbf{1.00}
    & 0.03\% & 0.03\% & \textbf{0.00\%}
    & \textbf{8.2}  & \textbf{8.4} & 18.2 \\ \hline
    & \textbf{Avg}
    & 0.71 & 0.77 & \textbf{0.96}
    & 0.77 & 0.81 & \textbf{0.98}
    & 0.77\% & 0.17\% & \textbf{0.04\%}
    & 5.14 & \textbf{4.99} & 20.20\\
\end{tblr} }
    \label{tab:traditional}
\end{table*}

\noindent \textbf{Attack Configurations.}
We evaluate \ac{qpa}-based \ac{sdm} with six non-adaptive query-based black-box attacks and deploy query-blinding and \ac{oars} strategies for these non-adaptive attacks to evaluate the robustness of our system against adaptive attacks. 

We selected six non-adaptive attacks: NES~\cite{ilyas2018black}, Boundary~\cite{brendel2018decisionbased}, Square~\cite{andriushchenko2020square}, HSJA~\cite{chen2020hopskipjumpattack}, QEBA~\cite{li2020qeba} and Surfree~\cite{maho2021surfree}. NES and Square are score-based attacks that need to get a full probability distribution across labels and the others are decision-based that only require the output label. We use the open-source implementation of the attacks and the default configurations. We set the normalized perturbation budgets $\sigma = 0.05$ as prior work used. The only exception is we set $\sigma = 0.1$ for the attacks on MNIST since it is necessary for them to succeed within a limited time.  Table~\ref{tab:NonAdaptiveAttack} lists the configurations and the attack performance of the non-adaptive attacks. For each configuration, we randomly select 100 images from the corresponding dataset. We run an attack on each instance until it terminates within 100K queries. As shown in Table~\ref{tab:NonAdaptiveAttack}, these attacks can achieve success within 50K queries for each instance. However, some instances still failed to generate adversarial examples even after 100K queries and we regard them as failed attacks.

For the adaptive attacks, we deploy two strategies on the above six non-adaptive attacks:  query-blinding and \ac{oars}. 
For query-blinding, we use standard attack configurations and deploy the standard random affine transformation to blind the queries by rotation, shifting and scaling. We set the rotation angle up to 10 degrees, shifting up to 10\% horizontally/vertically, and zooming up to 10\%. 
For \ac{oars}, it modifies the standard attack algorithms with an adapt-and-resample strategy. We use the open-source implementation and the default configurations of \ac{oars}. Due to \ac{oars} utilizing the leaked information of the decision of the defense model to adjust the attack direction and step size, we integrate our \ac{qpa}-based \ac{sdm} behind it. This integration allows us to return the decision for each input query throughout the entire attack procedure.

\noindent \textbf{\ac{qpa} Configurations.} By default, we employ the feature extraction algorithm of PIHA due to its effectiveness and efficiency. However, the feature extraction algorithm of PIHA is specifically designed for the three-channel images. When applied to the single-channel images of the MNIST dataset, we attempted to superimpose single-channel images into three-channel images but the results were suboptimal. Therefore, we use the feature extraction algorithm of BlackLight for the MNIST dataset. We preserve top $K = 20$ query provenance graphs in the memory and conduct detection on the graphs with more than $s = 15$ nodes. 
The setting of our hyperparameters will be discussed in Section~\ref{sec:hyperparameters}. When deploying \ac{qpa} on different datasets, the hyperparameters should be set following the method of our experiment in Section~\ref{sec:hyperparameters}.

\noindent \textbf{Real-world Scenarios.} In real-world scenarios, benign queries are often much more than attack queries, making detection more challenging. To simulate a less adversarial environment, we set the anomaly rate of the input sequence to 1\%~\cite{10177782}. Attackers may employ collaborative attacks to generate adversarial examples, and benign queries may exhibit high similarity when observing the same object from different angles or under varying lighting conditions. We will evaluate the performance of \ac{qpa} under these real-world conditions in Section~\ref{sec:realworld}.

\begin{table}[t!]
    \scriptsize
    \centering
    \caption{The \ac{asr} of the attacks with two adaptive strategies under the defense of \ac{qpa} and two baselines. The best results are in \textbf{bold}.}
    \resizebox{0.48\textwidth}{!}{\begin{tblr}{      
        cell{1-2}{1-8} = {c},
        cell{3-27}{1-2} = {l},
        cell{3-27}{3-8} = {r},
        hline{1,28} = {1pt}
        }
    \SetCell[r = 2, c = 1]{c} \textbf{Dataset}  & \SetCell[r = 2, c = 1]{c} \textbf{Attack} & \SetCell[r = 1, c = 3]{c} \textbf{Qurey Blinding}& &  & \SetCell[r = 1, c = 3]{c} \textbf{OARS} & &\\  \cline{3-8}
    & & \SetCell[r = 1, c = 1]{c}\textbf{BL} & \SetCell[r = 1, c = 1]{c}\textbf{PIHA} & \SetCell[r = 1, c = 1]{c}\textbf{\ac{qpa}} & \SetCell[r = 1, c = 1]{c}\textbf{BL} & \SetCell[r = 1, c = 1]{c}\textbf{PIHA} & \SetCell[r = 1, c = 1]{c}\textbf{\ac{qpa}}  \\\hline
    \SetCell[r = 6,c = 1]{r} \textbf{MNIST} &   NES & \textbf{0\%} & - & \textbf{0\%} &  2\%  &  - & \textbf{0\%}    \\\hline 
    & Boundary & \textbf{0\%} & -& \textbf{0\%} & 69\%  & - & \textbf{8\%}                             \\ \hline
    & Square & \textbf{0\%} &- & \textbf{0\%} & 55\%  & - & \textbf{7\%}                            \\ \hline
    & HSJA   & \textbf{0\%} &-& \textbf{0\%} & 76\%  & - &  \textbf{10\%}                          \\ \hline
    & QEBA  & \textbf{0\%} &- & \textbf{0\%} & 87\%  & - &   \textbf{6\%}                           \\ \hline
    & SurFree & 7\% & - & \textbf{0\%} & 71\%  & - & \textbf{0\%}  \\ \hline
    \SetCell[r = 6]{r}  \textbf{CIFAR10}   &  NES  & \textbf{0\%} & \textbf{0\%} & \textbf{0\%} &  99\% & 83\% & \textbf{0\%}   \\ \hline
    & Boundary  & \textbf{0\%} & \textbf{0\%} & \textbf{0\%} & 98\%  & 90\% & \textbf{0\%}                         \\ \hline
    & Square  & \textbf{0\%} & \textbf{0\%} & \textbf{0\%} & 93\%  & 99\% & \textbf{20\%}                       \\ \hline
    & HSJA   & \textbf{0\%} & \textbf{0\%} & \textbf{0\%} &  82\% & 76\% & \textbf{5\%}                      \\ \hline
    & QEBA   & \textbf{0\%}& \textbf{0\%} &\textbf{0\%} &  98\% &  95\%& \textbf{18\%}                           \\ \hline
    & SurFree & 5\% & 4\% & \textbf{2\%} & 81\%  & 67\% & \textbf{6\%}                         \\ \hline
    \SetCell[r = 6]{r}  \textbf{ImageNet}  & NES & 3\% & \textbf{0\%} & \textbf{0\%} &  100\%    & 92\% & \textbf{0\%}    \\\hline
    & Boundary  & 27\% & 18\% & \textbf{0\%} & 38\%  & 46\% & \textbf{0\%}                               \\ \hline
    & Square  & 1\% & \textbf{0\%} & \textbf{0\%} & 84\%  & 87\% & \textbf{0\%}                           \\ \hline
    & HSJA  & \textbf{0\%} &\textbf{0\%}& \textbf{0\%} & 50\%  & 91\% & \textbf{0\%}                             \\ \hline
    & QEBA  & \textbf{0\%} & \textbf{0\%} & \textbf{0\%} &  50\% & 96\% & \textbf{0\%}                            \\ \hline
    & SurFree & 2\% & \textbf{0\%} & \textbf{0\%} & 93\%  & 100\% & \textbf{0\%}                              \\ \hline
    \SetCell[r = 6]{r}  \textbf{CelebaHQ}  & NES & 40\% & 47\% & \textbf{0\%} &  100\% & 97\% & \textbf{8\%}    \\\hline
    & Boundary  & 96\% & 54\% &\textbf{0\%} & 73\%  & 70\% & \textbf{0\%}                               \\ \hline
    & Square  & 6\% & 9\% & \textbf{5\%} & 96\%  & 100\% & \textbf{0\%}                           \\ \hline
    & HSJA & \textbf{0\%} & \textbf{0\%} & \textbf{0\%} & 77\%  & 90\% & \textbf{3\%}                             \\ \hline
    & QEBA  & \textbf{0\%} & \textbf{0\%} & \textbf{0\%} &  93\% & 100\% & \textbf{8\%}                            \\ \hline
    & SurFree & \textbf{0\%} & 7\% & \textbf{0\%} & 98\%  & 100\% & \textbf{0\%}                              \\ \hline
    & \textbf{Avg} & 7.79\% & 7.72\% & \textbf{0.29\%}  & 77.63\% & 87.72\% & \textbf{4.08\%} \\ 
\end{tblr} }
    \label{tab:oars}
\end{table}

\subsection{Effectiveness} \label{sec:effectiveness}
\noindent \textbf{Non-adaptive Attacks.} We evaluate the detection performance of \ac{qpa} on the non-adaptive attacks in the simulated real-world environment. We set the anomaly rate of the input sequence as 1\% and limited the maximum number of queries to 50K. 
Our two baselines have shown promising performance on the non-adaptive attacks and they can make these non-adaptive attack algorithms achieve a 0\% attack success rate.  
We compare the detection coverage, detection precision, \ac{fpr}, and the \ac{ttd} of \ac{qpa} with BlackLight and PIHA. Table~\ref{tab:traditional} shows the detection performance of \ac{qpa} on the non-adaptive attacks. For each row in the table, we randomly sample 100 instances from the dataset and calculate the average number.

Overall, \ac{qpa} achieves the best performance on detection coverage, detection precision and \ac{fpr} on average. Our system achieves an average of  96\% detection coverage, 98\% detection precision and 0.04\% \ac{fpr}, maintaining similar results across various datasets and attack algorithms. Conversely, BlackLight and PIHA exhibit significant performance fluctuations, particularly in detecting Boundary and SurFree attacks. This is attributed to their excessively high thresholds for Boundary and SurFree, highlighting the unstable performance of individual similarity-based \ac{sdm}.
While the average number of queries to detection for \ac{qpa} is 20.20, slightly higher than BlackLight and PIHA, existing attacks necessitate hundreds to thousands of queries for a successful attack, as outlined in Table~\ref{tab:NonAdaptiveAttack}. 
Therefore, the results demonstrate that \ac{qpa} can effectively detect non-adaptive attacks with high efficiency and robustness.

\noindent \textbf{Adaptive Attacks.} We evaluate the detection performance of \ac{qpa} on the adaptive attacks with two strategies: query-blinding and \ac{oars}. Because \ac{oars} leverages the leaked information of the decision of the defense model to adjust the attack direction and step size, we integrate \ac{qpa} and our two baselines for the defense to return the decision for each input query while attacking. We record the \ac{asr} to show their defense performance. Table~\ref{tab:oars} shows the detection performance of \ac{qpa} and our two baselines on the adaptive attacks. We randomly sample 100 instances from the corresponding dataset for each attack configuration.

For the query-blinding strategy, our two baselines demonstrate substantial defense performance against adaptive attacks. However, \ac{qpa} outperforms them, achieving an average \ac{asr} of 0.29\%. BlackLight and PIHA exhibit poor defense performance against the Boundary attack, aligning with the results of non-adaptive attacks. 
For the \ac{oars} strategy, \ac{qpa} can reduce the \ac{asr} of the adaptive attacks to 4.08\% on average, which is 20.26$\times$ lower than the defense performance of the two baselines. \ac{qpa} achieves the best defense performance across all the attack algorithms and datasets. The attacks always terminate within 100K queries because they cannot find any available direction and step size to take the next step. \ac{oars} is designed to adapt the attack direction and step size based on the decision of the defense model to reduce the detection rate of the defense. However, \ac{qpa} can effectively detect each attempt and make the attack fail. 
The \ac{asr}s of Square and QEBA on CIFAR10 are the highest, at 20\% and 18\% respectively. This is because these attack algorithms perform extremely well on some samples and can generate adversarial examples within ten queries, which is hard for the defense to detect.

\subsection{Efficiency}\label{sec:efficiency}
\begin{figure}[t!]
    \centering
    \includegraphics[width=0.40\textwidth]{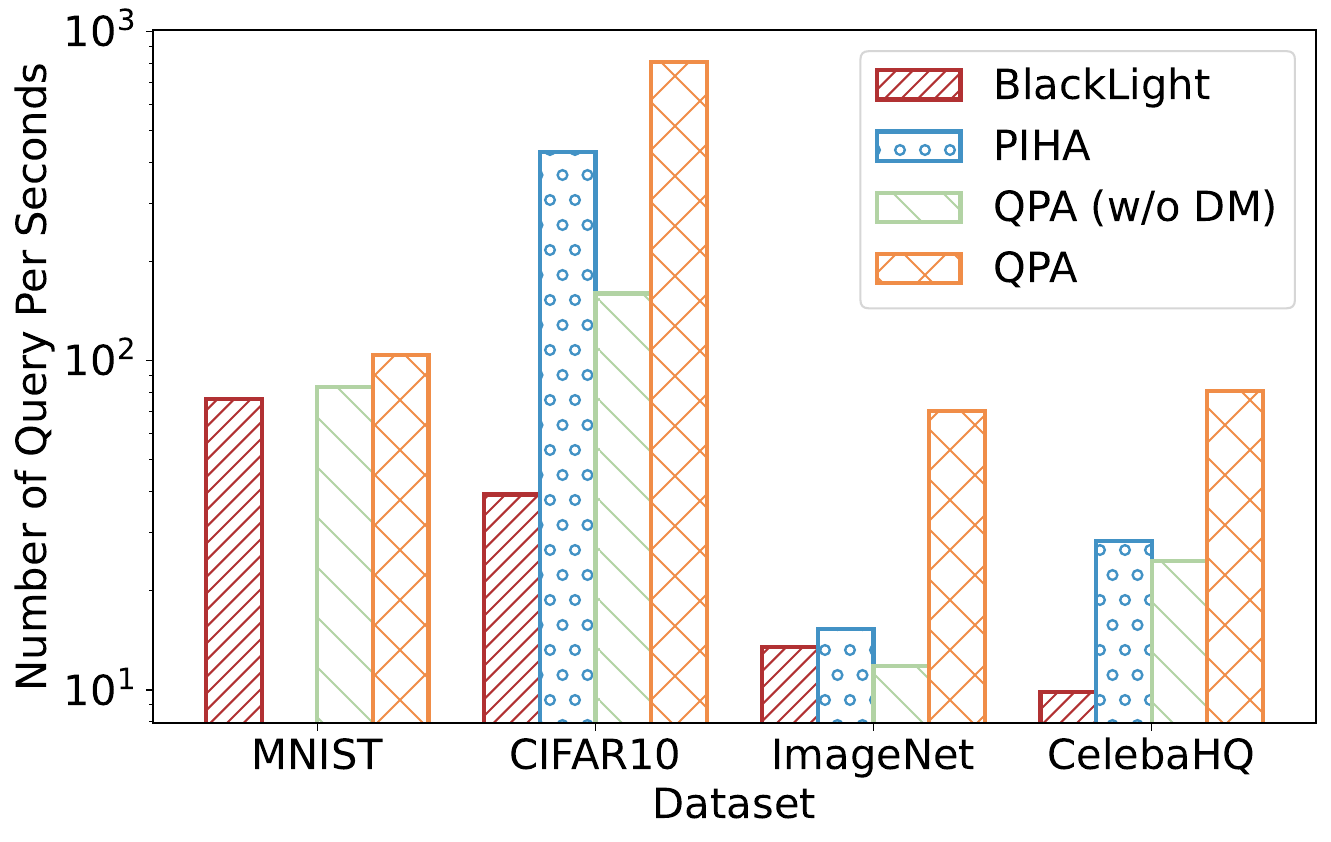}
    \caption{Throughput of \ac{qpa} compared with BlackLight, PIHA and \ac{qpa} without dynamic management. The throughput is the number of queries processed per second.}
    \label{fig:throughput}
\end{figure}

\begin{figure}[t!]
    \centering
    \includegraphics[width=0.48\textwidth]{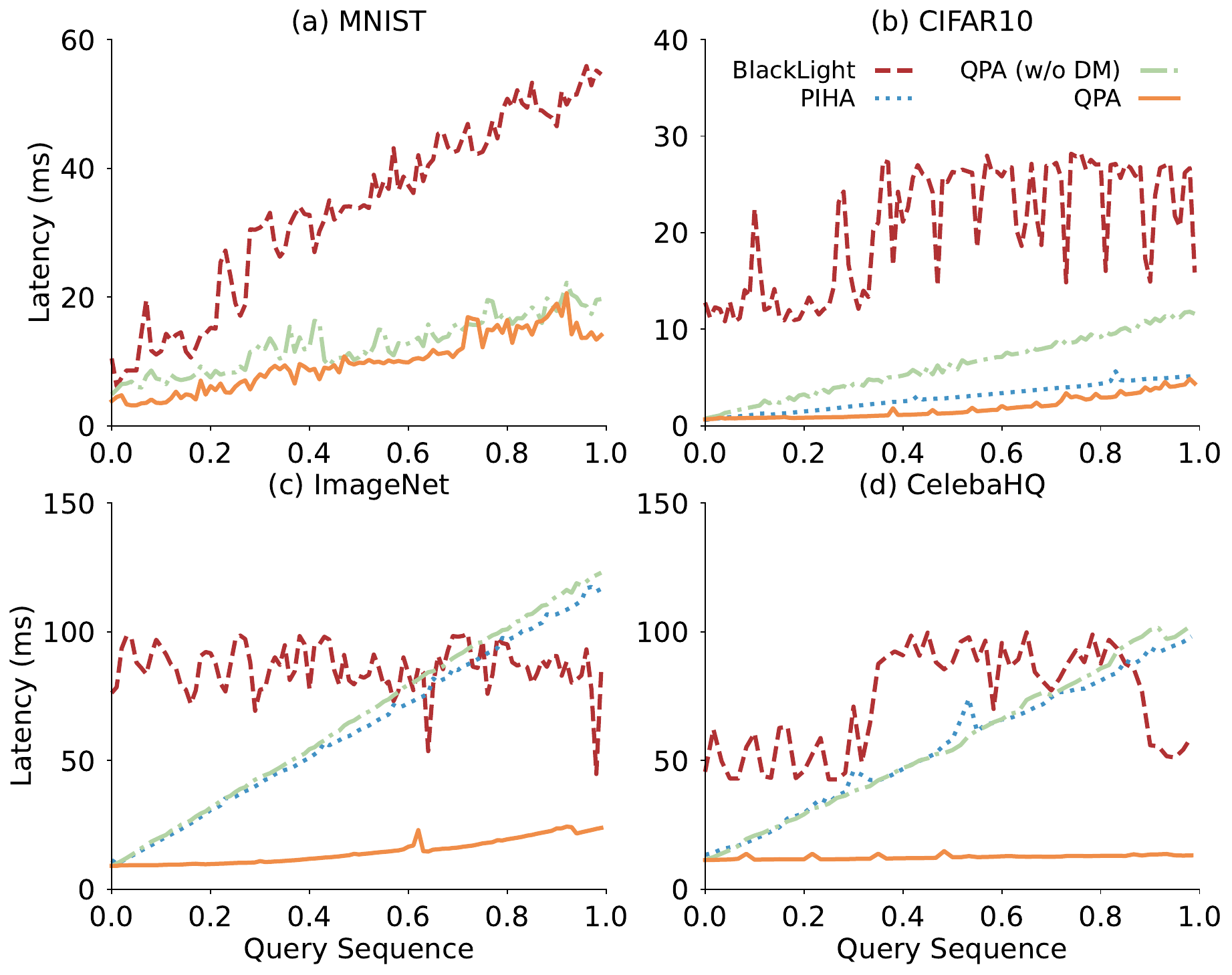}
    \caption{Latency of \ac{qpa} compared with BlackLight, PIHA and \ac{qpa} without dynamic management. The latency is the average time taken to process a query. The x-axis represents the proportion of the total sequence and the y-axis represents the response latency. Each point in the graph represents the average latency of 500 queries.}
    \label{fig:latency}
\end{figure}

We evaluate the efficiency of \ac{qpa} by comparing the throughput and latency with our baselines and our system without dynamic management.  We record the total processing time for each instance sequence and the latency of each query during the experiments in Section~\ref{sec:effectiveness}. We calculate the average throughput and latency for each system. We show our results in Figure~\ref{fig:throughput} and Figure~\ref{fig:latency}.

As shown in Figure~\ref{fig:throughput}, \ac{qpa} achieves the highest throughput among the three systems. The throughput of \ac{qpa} is 7.67$\times$  higher than BlackLight and 2.25$\times$ higher than PIHA. The larger the image, the more significant the throughput improvement. 
In Figure~\ref{fig:latency}, \ac{qpa} has the lowest latency among the three systems. The average latency of \ac{qpa} is 9.87ms for MNIST, 1.86ms for CIFAR10, 14.51ms for ImageNet and 12.45ms for CelebaHQ, which is 3.39 - 11.09$\times$ faster than BlackLight and 1.59 - 4.49$\times$ faster than PIHA. The dynamic management of our system improves the throughput by 3.81$\times$ and the latency by 1.29 - 4.56$\times$ compared with the system without dynamic management. The throughput and latency on ImageNet and CelebaHQ are slightly higher than those on MNIST and CIFAR10 due to the larger image size.
The metrics of CIFAR10 outperform MNIST because we use the similarity algorithm of BlackLight on MNIST, which is less efficient than the similarity algorithm of PIHA that we use in other datasets. The results on the other three datasets show that our system is more efficient than PIHA. The reason our system performs better than the other two baselines is that \ac{qpa} deploys graph eviction strategy to reduce the features stored in the memory so that it reduces the time to find the closest query in history.

\begin{figure}[t!]
    \centering
    \includegraphics[width=0.49\textwidth]{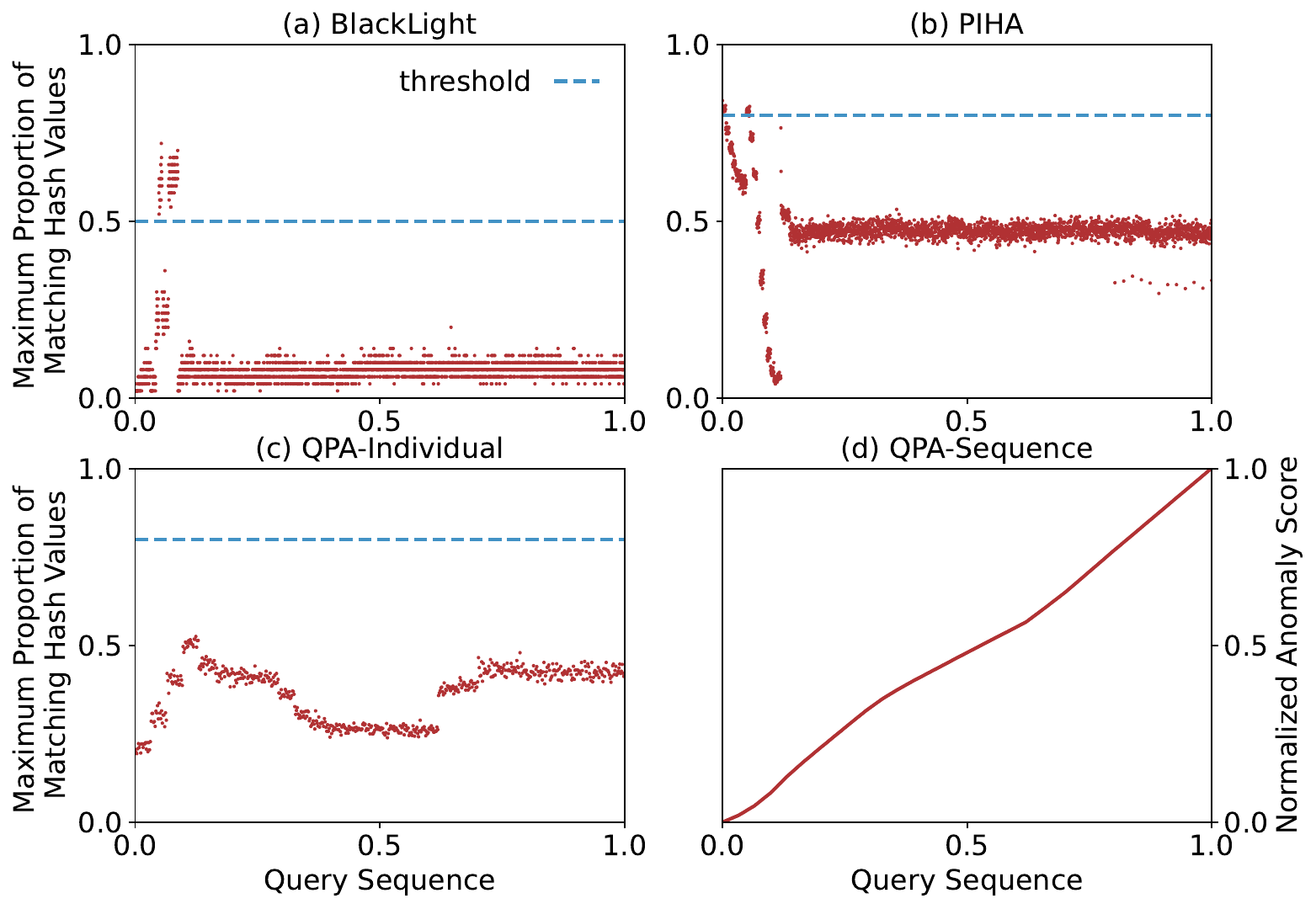}
    \caption{Comparison of the individual feature in existing \ac{sdm} with the sequence feature of our system in detecting NES-OARS. The x-axis represents the number of queries and the y-axis represents the maximum proportion of matched hash values for the individual feature distance and normalized anomaly score for sequence feature distance. (a) and (b) are the results of the two baselines, (c) is the individual features of our system and (d) is the sequence feature of our system. The blue dashed line represents the detection threshold.}
    \label{fig:distance}
\end{figure}

\begin{figure*}[!t]
    \centering 
    \includegraphics[width=0.98\textwidth]{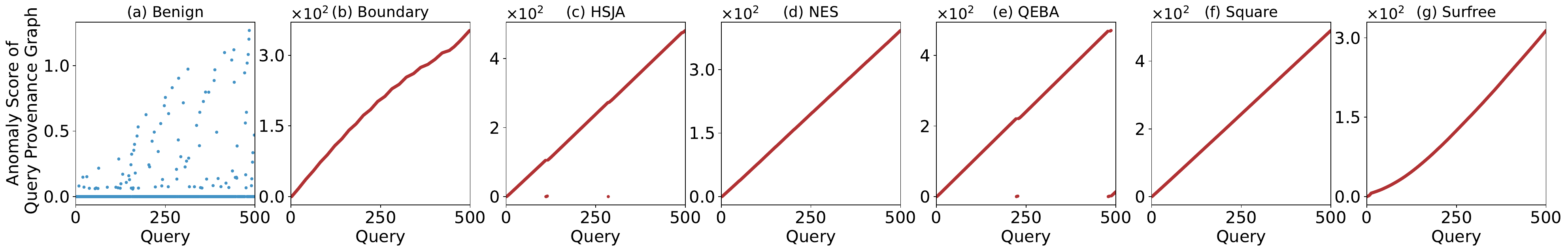}
    \caption{The statistics distribution of the query provenance graphs' anomaly score collected during the attacks with \ac{oars} strategy. We show the anomaly score of each query in the first 500 queries of benign and six attack query sequences. The x-axis denotes the query number, and the y-axis represents the anomaly score of the query provenance graph for each query.}
    \label{fig:distribution}
\end{figure*}

\subsection{Insight Verification}\label{sec:insightverification}

While the \ac{qpa} has demonstrated impressive performance in detecting both non-adaptive and adaptive attacks, it remains crucial to validate the key insight that sequence features are more robust than individual features. To this end, we have conducted the following experiments. 

\noindent \textbf{Individual Feature vs. Sequence Feature.}
We first compare the statistics of individual features and sequence features in detecting NES attack with \ac{oars} strategy, denoted as NES-OARS, on ImageNet. We deploy \ac{qpa} and the other two baselines to defend against the NES-OARS attacks, as configured in the adaptive attacks experiments. We record the individual features for each query during the attack procedure. For \ac{qpa}, we also collect the normalized anomaly score of the query provenance graph to which the new query is appended as the sequence feature. 

Figure~\ref{fig:distance} demonstrates that the sequence feature of \ac{qpa} is more robust than individual features in detecting adaptive attacks. The NES-OARS attack successfully evaded BlackLight and PIHA after 3,075 and 2,839 queries but failed to evade QPA after 614 query attempts due to the lack of feasible directions. For individual similarity-based detection, 97.40\% of the queries evade BlackLight's detection and 98.66\% PIHA's. Although all individual features collected during QPA's detection are below the detection threshold, QPA's sequence feature still successfully detects the attack. As the attack proceeds, the sequence feature of \ac{qpa} gradually increases. This indicates that while adaptive attacks can adjust their perturbations to evade the threshold, they struggle to conceal the sequence feature. These results confirm that sequence features can capture the attack pattern more effectively than individual features.

\begin{figure}[t!]
    \centering
    \includegraphics[width=0.45\textwidth]{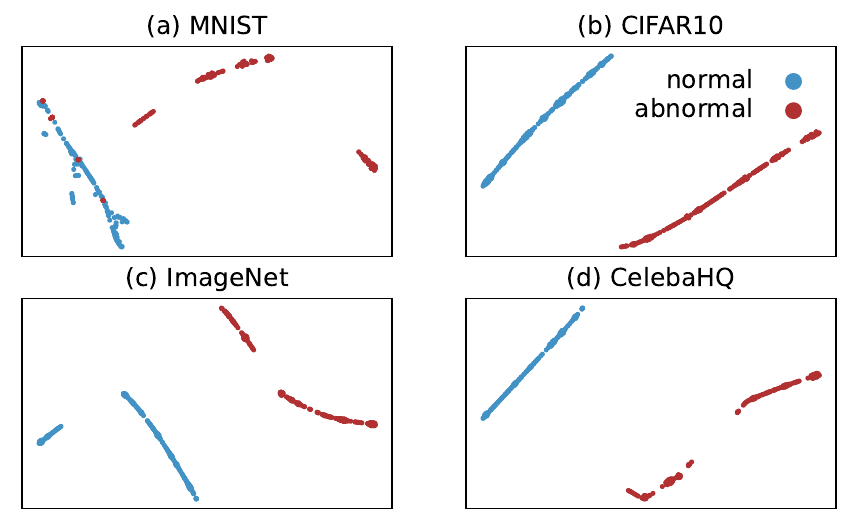}
    \caption{T-SNE visualization of the graph embedding. The blue points represent the benign query provenance graphs and the red points represent the anomaly ones. The visualization results show that the graph embedding can effectively distinguish the anomaly query provenance graphs from the benign query provenance graphs.}
    \label{fig:embedding}
\end{figure}

\noindent \textbf{Statistics Distribution.} To verify the distinctiveness of sequence features from individual features, we analyzed the anomaly scores of query provenance graphs, comparing sequences of benign and adversarial queries. In our experiments, we randomly sampled an instance from the ImageNet as the victim for each type of attack using \ac{oars} strategy, subsequently recording the anomaly score of the query provenance graph to which the new query is appended during the attack procedure. Given that attack sequences encompass hundreds to thousands of queries, we present the results pertaining to the first five hundred queries in Figure~\ref{fig:distribution}. 
Our analysis reveals that the query provenance features of benign queries are magnitudes smaller than those of the attack queries, with a substantial 84\% of benign queries exhibiting no discernible query provenance. This absence of query provenance among benign queries is attributable to their lack of inter-relatedness, resulting in a query provenance graph that reflects random, disparate behavior. Conversely, the anomaly score for adversarial queries escalates as successive queries arrive, a trend that is a direct consequence of their consecutive generation. This analysis underscores the statistical distinctiveness of sequence features from individual features, providing a basis for statistics analysis to detect adaptive attacks.

\noindent \textbf{Graph Embedding Visualization.} To validate the effectiveness of the graph classifier in distinguishing sequence features from individual features, we visualized the graph embeddings of query provenance graphs. We collect the outputs of statistics analysis from 100 instances in the experiments of NES-OARS for each dataset in Section~\ref{sec:effectiveness}. Using the pre-trained graph classifier, we extracted the graph embeddings of these query provenance graphs. We then use the t-SNE algorithm for visualization. As shown in Figure~\ref{fig:embedding}, the query provenance graph of the benign queries is clearly delineated from the anomalous query provenance graph. This visualization underscores the efficacy of graph embeddings in differentiating benign query provenance graphs from anomalous ones, indicating the robustness of sequence features in detecting adaptive attacks.

\subsection{Hyperparameters and Ablation Study}\label{sec:hyperparameters}

\begin{figure}[!t]
    \centering
    \includegraphics[width=0.49\textwidth]{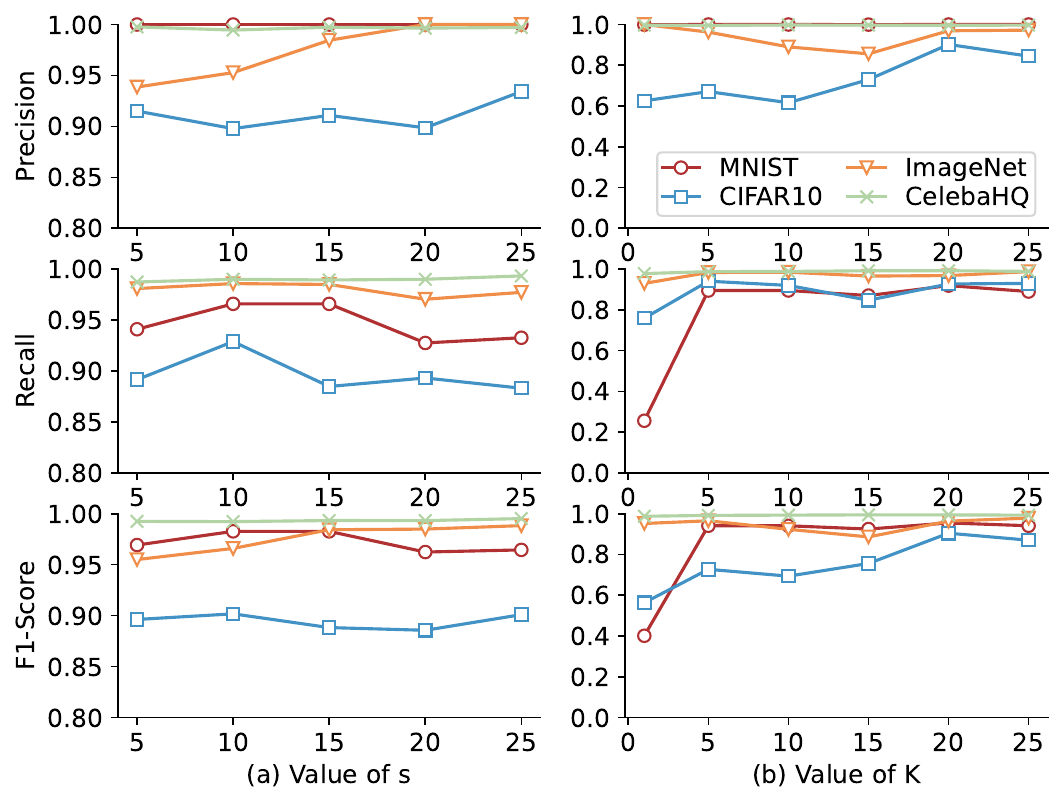}
    \caption{Detection performance of our system on different parameter settings.}
    \label{fig:para-ttd}
\end{figure}

\noindent \textbf{Hyperparameters Setting.}
We evaluate the setting of hyperparameters in \ac{qpa} to show the impact of each hyperparameter on the detection performance. We choose the default hyperparameters using the optimal experiment results on the four datasets with the HSJA attack. We record the detection precision, recall and F1-Score as metrics to evaluate the detection performance. For each setting, we run the experiments 20 times and record the average results. We show the results in Figure~\ref{fig:para-ttd}.

\noindent \textbf{\textit{Minimum graph size $s$ to detect.}}
In \ac{qpa}, we set a minimum graph size $s$ to limit the input size to the graph classifier. The larger the $s$ setting, the longer the \ac{ttd}. Therefore we evaluate the impact of $s$ on detection performance so that we can balance it with \ac{ttd}. As shown in Figure~\ref{fig:para-ttd}(a), the detection precision increases with the increase of $s$. The reason is that as the graph size increases, the graph classifier is more robust to capture the attack pattern. However, the detection recall decreases when $s$ is too large. That is because the graph classifier may filter out some small graphs containing the attack pattern. We set $s = 15$ for it achieves the best average F1-Score on our datasets.

\noindent \textbf{\textit{Top $K$ graphs.}}
We use the parameter $K$ to control the number of graphs stored in the cache. The larger the $K$ setting, the more features will be stored in the memory, which will consume more memory.   
We evaluate the impact of $K$ on the detection performance. As shown in Figure~\ref{fig:para-ttd}(b), the detection performance increases as the $K$ increases and achieves the best performance when $K = 20$. The reason is that the more graphs stored in the cache, the more features can be used to detect the attack pattern. However, the detection efficiency decreases when $K$ increases. We set $K = 20$ because it achieves the best average F1-Score on the four datasets.

\begin{table}[t!]
    \scriptsize
    \centering
    \caption{Ablation Study on the detection performance of our system. ``w/o Threshold'' removes the threshold mechanism in the graph construction. ``w/o Graph Classifier'' removes the graph classifier in the \ac{qpa}. We regard the detection performance of the original system as the baseline. ``Diff'' shows the decrease in the detection performance compared with the baseline.}
    \resizebox{0.48\textwidth}{!}{\begin{tblr}{      
        cell{1-2}{1-7} = {c},
        cell{3-8}{2-7} = {r},
        cell{3-8}{1} = {l},
        hline{1,9} = {1pt}
        }
    \SetCell[r = 2, c = 1]{c} \textbf{Dataset}  & \SetCell[r = 1, c = 3]{c} \textbf{w/o Threshold} & & & \SetCell[r = 1, c = 3]{c} \textbf{w/o Graph Classifier}& &  \\  \cline{2-7}
    & \SetCell[r = 1, c = 1]{c}\textbf{Precsion} & \SetCell[r = 1, c = 1]{c}\textbf{Recall} & \SetCell[r = 1, c = 1]{c}\textbf{FPR} & \SetCell[r = 1, c = 1]{c}\textbf{Precision} & \SetCell[r = 1, c = 1]{c}\textbf{Recall} & \SetCell[r = 1, c = 1]{c}\textbf{FPR}  \\\hline
    \SetCell[r = 1,c = 1]{r} \textbf{MNIST} & 0.25 & 0.24 & 0.75\% & 0.24 & 0.86 & 4.74\% \\ \hline
    \SetCell[r = 1, c = 1]{r}  \textbf{CIFAR10} & 0.31 & 0.79 & 1.79\% & 0.09 & 0.60 & 9.26\% \\ \hline
    \SetCell[r = 1, c = 1]{r}  \textbf{ImageNet} & NA & NA & NA & 0.39 & 0.97 & 3.19\% \\ \hline
    \SetCell[r = 1, c = 1]{r}  \textbf{CelebaHQ} & 0.31  & 0.65 & 1.38\% & 0.77 & 0.99 & 3.98\% \\ \hline
    \SetCell[r = 1, c = 1]{r}  \textbf{Avg} & 0.29  & 0.56 & 1.31\% & 0.37 & 0.86 & 5.29\% \\ \hline
    \SetCell[r = 1, c = 1]{r} \textbf{Diff} & -70.41\%  & -42.86\% & $\infty$ & -61.99\% & -12.76\% & $\infty$ \\
\end{tblr} }
    \label{tab:ablation}
\end{table}

\noindent \textbf{Ablation Study.}
Our system incorporates a threshold mechanism to filter benign queries and a graph classifier to enhance detection accuracy. We conducted an ablation study on four datasets under HSJA attacks to evaluate the individual contributions of these components. This study involved removing the threshold mechanism from the graph construction phase and the graph classifier from the \ac{qpa} phase. We assessed the impact of each modification on detection performance separately, using the performance metrics of the unmodified system as a baseline. Each experimental configuration was tested 100 times to ensure statistical robustness, with the average results being reported for analysis.

Table \ref{tab:ablation} reveals a substantial decline in detection performance with the removal of the threshold mechanism and the graph classifier. The average precision for detection falls by 70.41\% and 61.99\% respectively. Correspondingly, the average recall for detection decreases by 42.86\% and 12.76\%. Additionally, the false positive rate (\ac{fpr}) increases from 0\% to 1.31\% and 5.29\%.
The system encounters complete failure in detection on the ImageNet dataset without the threshold mechanism. This highlights its crucial role in preventing benign queries from connecting to anomalous graphs. Without it, the graph structure and anomaly score distribution are significantly altered, hindering the effectiveness of the \ac{qpa}. The absence of a graph classifier also compromises the system's ability to analyze graph structures and identify attack patterns, leading to an increase in false positives. These results emphasize the essential role of both the threshold mechanism and the graph classifier in achieving optimal detection performance.

\subsection{Detection in Real-world Scenarios}\label{sec:realworld}
Although the experiments in Section~\ref{sec:effectiveness} have demonstrated the effectiveness of \ac{qpa} in detecting non-adaptive and adaptive attacks, the detection in real-world settings is more complex and challenging. To evaluate the performance of \ac{qpa} in real-world settings, we conduct two experiments: the detection of the collaborative attacks and similar benign queries.

\noindent \textbf{Collaborative Attacks.}
In real-world attacks, attackers can use different methods simultaneously, which may result in mixed query graphs and potentially lead to poor detection results. Conceptually, the sequences of different attack methods can each form a unique query provenance graph (targeting different victim queries) or a united query provenance graph (targeting the same victim query). These query provenance graphs are distinguishable from the benign ones so that \ac{qpa} can defend against them robustly.

To evaluate the detection performance of \ac{qpa} in detecting collaborative attacks, we randomly selected four attacks from a pool of twelve, which included six non-adaptive attacks and their corresponding \ac{oars} versions. These attacks were used to generate adversarial examples on both the same and different victim images. We mixed the first 200 queries of each attack algorithm with normal queries, maintaining an anomaly rate of 1\% to simulate a low anomaly rate in a real-world scenario. We repeat this test ten times for each dataset. Table~\ref{tab:colleborated} presents the detection performance of \ac{qpa} on the collaborative attacks. The results indicate that \ac{qpa} achieves an average recall of 0.97, an average precision of 0.94 and an average \ac{fpr} of 0.06\% when attacks target the same victim. When attacks are on different victims, \ac{qpa} achieves an average recall of 0.94, an average precision of 0.93 and an average \ac{fpr} of 0.07\%. The results demonstrate that \ac{qpa} can robustly detect collaborative attacks.

\begin{table}[t!]
    \scriptsize
    \centering
    \caption{The detection performance of \ac{qpa} on collaborative attacks. ``Same Victim'' represents the detection performance when the attacks target the same victim. ``Different Victims'' represents the detection performance when the attacks target different victims.}
    \resizebox{0.48\textwidth}{!}{\begin{tblr}{      
        cell{1-2}{1-4} = {c},
        cell{3-7}{1} = {l},
        cell{3-7}{2-7} = {r},
        hline{1,8} = {1pt}
        }
    \SetCell[r = 2, c = 1]{c} \textbf{Dataset}  & \SetCell[r = 1, c = 3]{c} \textbf{Same Victim} & & & \SetCell[r = 1, c = 3]{c} \textbf{Different Victims} & &  \\  \cline{2-7}
    & \SetCell[r = 1, c = 1]{c} \textbf{Precision} & \SetCell[r = 1, c = 1]{c} \textbf{Recall} & \SetCell[r = 1, c = 1]{c} \textbf{FPR} & \SetCell[r = 1, c = 1]{c} \textbf{Precision} & \SetCell[r = 1, c = 1]{c} \textbf{Recall} & \SetCell[r = 1, c = 1]{c} \textbf{FPR} \\ \hline
    \SetCell[r = 1,c = 1]{r} \textbf{MNIST} & 0.93 & 0.95 & 0.07\% & 0.91 & 0.93 & 0.09\%\\\hline 
    \SetCell[r = 1]{r}  \textbf{CIFAR10}  & 0.91 & 0.96 &  0.09\%  & 0.89  & 0.92 & 0.11\% \\\hline 
    \SetCell[r = 1]{r}  \textbf{ImageNet} & 0.96 & 0.99 & 0.04\%  & 0.95 & 0.97 & 0.05\% \\\hline 
    \SetCell[r = 1]{r}  \textbf{CelebaHQ} & 0.97 & 0.99 & 0.03\% & 0.98 & 0.95 & 0.02\% \\ \hline
    \SetCell[r = 1]{r}  \textbf{Avg} & 0.94  & 0.97 & 0.06\% & 0.93 &  0.94 &  0.07\% \\
\end{tblr} }
    \label{tab:colleborated}
\end{table}

\noindent \textbf{Similar Benign Queries.}
In real-world scenarios, benign images may exhibit high similarity with each other. For example, the same object could be captured at different times and under varying observation angles, lighting conditions, and other factors. These benign queries may have similar features, potentially leading to false positives.

We evaluated the \ac{fpr} using the YouTube Faces dataset~\cite{5995566}, which contains 3,425 videos of 1,595 individuals and is designed for face recognition learning. The frames within each video are highly similar, featuring the same subjects and backgrounds. We randomly selected 100 videos and treated each frame as a query, repeating this process ten times to calculate the \ac{fpr}. The \ac{fpr} of \ac{qpa} (0.23\%) is significantly lower than that of BlackLight (31.68\%) and PIHA (23.66\%). The feature extractor of \ac{qpa} smooths per-pixel noise, causing attack queries to form a dense cluster, while benign queries form multiple scattered clusters due to real-world variations (e.g., different observation angles). The high \ac{fpr} of the baselines is attributed to their tendency to match new queries with the most similar historical ones, resulting in high similarity scores when nearly identical benign queries are present (e.g., when the observation point remains static for a second). Therefore, \ac{qpa} can effectively reduce the \ac{fpr} in real-world settings.

\section{Discussion} \label{sec:discussion}

\noindent\textbf{Evasion from QPA.} There are two kinds of potential evasion attacks but both are ineffective against QPA. First, 
due to computational and storage constraints, \ac{qpa} periodically evicts benign query provenance graphs to disk and resets the query provenance graph storage. This could potentially allow adversaries to evade detection by launching a long-running attack: pausing when the attacker receives a rejection response, waiting for a long time, and resuming after the system reset.
Although this evasion attack is theoretically viable, the time required for a successful attack is very long and impractical for adversaries. As shown in Table~\ref{tab:NonAdaptiveAttack}, the average number of queries for an attack is 6,721. If we reset the database every time the adversary conducts average TTD queries in one reset period, a successful attack on one input sample would take an average of 333 days, which would be unrealistic for adversaries. 

Another potential approach is increasing the step size of query perturbations to reduce sequence aggregation. However, this would substantially lower the \ac{asr} of existing attacks~\cite{chen2020hopskipjumpattack,281294,maho2021surfree}. Experiments have shown that the \ac{asr} decreases as the step size increases, and the similarities between attack queries cannot be entirely prevented.

\noindent\textbf{Hyperparameters.} Hyperparameters can also influence the detection performance of \ac{qpa}. Our experiments revealed that \ac{qpa}'s performance is influenced by hyperparameters such as the minimum graph size $s$ required for detection and the Top $K$ graphs.
Considering the variability of datasets in real-world scenarios, it is essential to tune \ac{qpa}'s hyperparameters based on the specific deployment context. For optimal performance on particular datasets, defenders should adjust \ac{qpa}'s hyperparameters as demonstrated in Section~\ref{sec:hyperparameters}.

\noindent\textbf{Limitations.} Our \ac{qpa}-based \ac{sdm} encounter two limitations. 
First, the performance of \ac{qpa} relies on the effectiveness of feature extractors. The defender should properly choose the feature extractor based on the deployment scenario to distinguish attack and benign queries. 
Second, \ac{qpa}-based \ac{sdm} will occupy additional storage space (e.g., memory and disk storage) because we need to maintain a query provenance graph and store the information of graph edges. However, the number of edges is controllable and is equal to or less than the number of nodes (queries) because we only add one edge for each new query at most. Therefore the storage requirement remains $O(n)$, where $n$ is the number of queries.

\section{Other Related Work}
\noindent\textbf{Provenance Analysis.}
Provenance analysis has been widely studied in cyber security for Advanced Persistent Threat detection and investigation~\cite{king2003backtracking,HOLMES,nodlink,Han2020}. It leverages the historical relationships, modeled as provenance graph, between the system entities in system audit logs to detect malicious activities. Inspired by this, we propose to model query sequences as query provenance graphs to discover the relationships between queries and detect adversarial attacks.

\noindent\textbf{DNN Protection.}
Protecting the security and privacy of DNN models is an important topic in the security community. To defend against adversarial attacks, researchers have proposed various solutions such as adversarial training~\cite{shafahi2019adversarial} and state-based defenses~\cite{281294,10.1016/j.future.2023.04.005,zhang2020dynamic}. Researchers also focus on protecting DNN security in other important scenarios, such as transfer learning~\cite{zhang2022remos} and federated learning~\cite{zhang2023fedslice,liu2021distfl,liu2020pmc}. Recently, defenders tried to utilize trusted hardware (e.g., Intel SGX and ARM TrustZone) to protect the intellectual property of DNN and input privacy~\cite{zhang2023no}.

\section{Conclusion}
\label{sec:conclusion}
Query-based black-box attacks pose a significant threat to \ac{mlaas} systems. \ac{sdm} have been proposed to moderate the input queries based on their similarity to the history queries. However, these \ac{sdm} are not efficient enough for real-time detection and are vulnerable to sophisticated adaptive attacks, which can bypass the \ac{sdm} and achieve high \ac{asr}. To this end, 
we propose Query Provenance Analysis (QPA) for robust and efficient defense against query-based black-box attacks. 
We model the query sequences as query provenance graphs and design efficient algorithms with dynamic management of query provenance graphs to detect adversarial queries efficiently. 
Our experiments on existing \ac{sota} non-adaptive and adaptive attacks show that \ac{qpa} can significantly reduce the \ac{asr}, maintaining a low false positive rate and minimal overhead. Our work presents new challenges for future research on stronger black-box attack algorithms and adaptive strategies.

\section{Acknowledgment}

We thank the anonymous reviewers for their comments. This work was partly supported by the National Science and Technology Major Project of China (2022ZD0119103) and the National Natural Science Foundation of China (62172009)

\bibliographystyle{IEEEtranS}
\bibliography{references}

\begin{thebibliography}{10}
\providecommand{\url}[1]{#1}
\csname url@samestyle\endcsname
\providecommand{\newblock}{\relax}
\providecommand{\bibinfo}[2]{#2}
\providecommand{\BIBentrySTDinterwordspacing}{\spaceskip=0pt\relax}
\providecommand{\BIBentryALTinterwordstretchfactor}{4}
\providecommand{\BIBentryALTinterwordspacing}{\spaceskip=\fontdimen2\font plus
\BIBentryALTinterwordstretchfactor\fontdimen3\font minus
  \fontdimen4\font\relax}
\providecommand{\BIBforeignlanguage}[2]{{%
\expandafter\ifx\csname l@#1\endcsname\relax
\typeout{** WARNING: IEEEtranS.bst: No hyphenation pattern has been}%
\typeout{** loaded for the language `#1'. Using the pattern for}%
\typeout{** the default language instead.}%
\else
\language=\csname l@#1\endcsname
\fi
#2}}
\providecommand{\BIBdecl}{\relax}
\BIBdecl

\bibitem{Grubbs-test}
``Grubbs's test,'' 2022, https://en.wikipedia.org/wiki/Grubbs\%27s\_test.

\bibitem{amazonrekognition}
Amazon., ``Amazon rekognition: Automate your image recognition and video
  analysis with machine learning.'' 2024,
  https://aws.amazon.com/cn/rekognition/.

\bibitem{andriushchenko2020square}
M.~Andriushchenko, F.~Croce, N.~Flammarion, and M.~Hein, ``Square attack: a
  query-efficient black-box adversarial attack via random search,'' in
  \emph{European conference on computer vision}.\hskip 1em plus 0.5em minus
  0.4em\relax Springer, 2020, pp. 484--501.

\bibitem{brendel2018decisionbased}
W.~Brendel, J.~Rauber, and M.~Bethge, ``Decision-based adversarial attacks:
  Reliable attacks against black-box machine learning models,'' 2018.

\bibitem{chen2020hopskipjumpattack}
J.~Chen, M.~I. Jordan, and M.~J. Wainwright, ``Hopskipjumpattack: A
  query-efficient decision-based attack,'' in \emph{2020 IEEE Symposium on
  Security and Privacy (SP)}.\hskip 1em plus 0.5em minus 0.4em\relax IEEE,
  2020, pp. 1277--1294.

\bibitem{10.1145/3385003.3410925}
S.~Chen, N.~Carlini, and D.~Wagner, ``Stateful detection of black-box
  adversarial attacks,'' in \emph{Proceedings of the 1st ACM Workshop on
  Security and Privacy on Artificial Intelligence}, 2020, p. 30–39.

\bibitem{10.1016/j.future.2023.04.005}
S.-H. Choi, J.~Shin, and Y.-H. Choi, ``Piha: Detection method using perceptual
  image hashing against query-based adversarial attacks,'' \emph{Future Gener.
  Comput. Syst.}, vol. 145, no.~C, p. 563–577, 2023.

\bibitem{clarifai}
Clarifai., ``The world's ai: Clarifai computer vision ai and machine learning
  platform.'' 2024, https://www.clarifai.com/.

\bibitem{6296535}
L.~Deng, ``The mnist database of handwritten digit images for machine learning
  research [best of the web],'' \emph{IEEE Signal Processing Magazine},
  vol.~29, no.~6, pp. 141--142, 2012.

\bibitem{9760120}
B.~Esmaeili, A.~Azmoodeh, A.~Dehghantanha, H.~Karimipour, B.~Zolfaghari, and
  M.~Hammoudeh, ``Iiot deep malware threat hunting: From adversarial example
  detection to adversarial scenario detection,'' \emph{IEEE Transactions on
  Industrial Informatics}, vol.~18, no.~12, pp. 8477--8486, 2022.

\bibitem{10.1145/3576915.3623116}
R.~Feng, A.~Hooda, N.~Mangaokar, K.~Fawaz, S.~Jha, and A.~Prakash, ``Stateful
  defenses for machine learning models are not yet secure against black-box
  attacks,'' in \emph{Proceedings of the 2023 ACM SIGSAC Conference on Computer
  and Communications Security (CCS)}, 2023, p. 786–800.

\bibitem{6248074}
A.~Geiger, P.~Lenz, and R.~Urtasun, ``Are we ready for autonomous driving? the
  kitti vision benchmark suite,'' in \emph{2012 IEEE Conference on Computer
  Vision and Pattern Recognition}, 2012, pp. 3354--3361.

\bibitem{Han2020}
X.~Han, T.~Pasquier, A.~Bates, J.~Mickens, and M.~Seltzer, ``Unicorn: Runtime
  provenance-based detector for advanced persistent threats,'' in
  \emph{Proceedings of the Network and Distributed System Security Symposium
  (NDSS)}, 2022.

\bibitem{ilyas2018black}
A.~Ilyas, L.~Engstrom, A.~Athalye, and J.~Lin, ``Black-box adversarial attacks
  with limited queries and information,'' in \emph{International conference on
  machine learning}.\hskip 1em plus 0.5em minus 0.4em\relax PMLR, 2018, pp.
  2137--2146.

\bibitem{juuti2019prada}
M.~Juuti, S.~Szyller, S.~Marchal, and N.~Asokan, ``Prada: protecting against
  dnn model stealing attacks,'' in \emph{2019 IEEE European Symposium on
  Security and Privacy (EuroS\&P)}.\hskip 1em plus 0.5em minus 0.4em\relax
  IEEE, 2019, pp. 512--527.

\bibitem{karras2017progressive}
T.~Karras, T.~Aila, S.~Laine, and J.~Lehtinen, ``Progressive growing of gans
  for improved quality, stability, and variation,'' \emph{arXiv preprint
  arXiv:1710.10196}, 2017.

\bibitem{king2003backtracking}
S.~T. King and P.~M. Chen, ``Backtracking intrusions,'' in \emph{Proceedings of
  the Nineteenth ACM Symposium on Operating Systems Principles (SOSP)}, 2003,
  p. 223–236.

\bibitem{krizhevsky2009learning}
A.~Krizhevsky, ``Learning multiple layers of features from tiny images,'' pp.
  32--33, 2009.

\bibitem{726791}
Y.~Lecun, L.~Bottou, Y.~Bengio, and P.~Haffner, ``Gradient-based learning
  applied to document recognition,'' \emph{Proceedings of the IEEE}, vol.~86,
  no.~11, pp. 2278--2324, 1998.

\bibitem{li2020qeba}
H.~Li, X.~Xu, X.~Zhang, S.~Yang, and B.~Li, ``Qeba: Query-efficient
  boundary-based blackbox attack,'' in \emph{Proceedings of the IEEE/CVF
  conference on computer vision and pattern recognition}, 2020, pp. 1221--1230.

\bibitem{281294}
H.~Li, S.~Shan, E.~Wenger, J.~Zhang, H.~Zheng, and B.~Y. Zhao, ``Blacklight:
  Scalable defense for neural networks against {Query-Based} {Black-Box}
  attacks,'' in \emph{31st USENIX Security Symposium (USENIX Security 22)},
  2022, pp. 2117--2134.

\bibitem{nodlink}
S.~Li, F.~Dong, X.~Xiao, H.~Wang, F.~Shao, J.~Chen, Y.~Guo, X.~Chen, and D.~Li,
  ``Nodlink: An online system for fine-grained apt attack detection and
  investigation.'' in \emph{NDSS}, 2024.

\bibitem{liu2021distfl}
B.~Liu, Y.~Cai, Z.~Zhang, Y.~Li, L.~Wang, D.~Li, Y.~Guo, and X.~Chen, ``Distfl:
  Distribution-aware federated learning for mobile scenarios,''
  \emph{Proceedings of the ACM on Interactive, Mobile, Wearable and Ubiquitous
  Technologies}, vol.~5, no.~4, pp. 1--26, 2021.

\bibitem{liu2020pmc}
B.~Liu, Y.~Li, Y.~Liu, Y.~Guo, and X.~Chen, ``Pmc: A privacy-preserving deep
  learning model customization framework for edge computing,''
  \emph{Proceedings of the ACM on Interactive, Mobile, Wearable and Ubiquitous
  Technologies}, vol.~4, no.~4, pp. 1--25, 2020.

\bibitem{liu2020computing}
L.~Liu, S.~Lu, R.~Zhong, B.~Wu, Y.~Yao, Q.~Zhang, and W.~Shi, ``Computing
  systems for autonomous driving: State of the art and challenges,'' \emph{IEEE
  Internet of Things Journal}, vol.~8, no.~8, pp. 6469--6486, 2020.

\bibitem{maho2021surfree}
T.~Maho, T.~Furon, and E.~Le~Merrer, ``Surfree: a fast surrogate-free black-box
  attack,'' in \emph{Proceedings of the IEEE/CVF Conference on Computer Vision
  and Pattern Recognition}, 2021, pp. 10\,430--10\,439.

\bibitem{HOLMES}
S.~M. Milajerdi, R.~Gjomemo, B.~Eshete, R.~Sekar, and V.~Venkatakrishnan,
  ``{HOLMES: real-time APT detection through correlation of suspicious
  information flows},'' in \emph{Proceedings of the IEEE Symposium on Security
  and Privacy (IEEE S\&P)}, 2019, pp. 1137--1152.

\bibitem{8835245}
M.~Nasr, R.~Shokri, and A.~Houmansadr, ``Comprehensive privacy analysis of deep
  learning: Passive and active white-box inference attacks against centralized
  and federated learning,'' in \emph{2019 IEEE Symposium on Security and
  Privacy (SP)}, 2019, pp. 739--753.

\bibitem{papernot2017practical}
N.~Papernot, P.~McDaniel, I.~Goodfellow, S.~Jha, Z.~B. Celik, and A.~Swami,
  ``Practical black-box attacks against machine learning,'' in
  \emph{Proceedings of the 2017 ACM on Asia Conference on Computer and
  Communications Security (ASIACCS)}, 2017, pp. 506--519.

\bibitem{10177782}
A.~Rashid and J.~Such, ``Malprotect: Stateful defense against adversarial query
  attacks in ml-based malware detection,'' \emph{IEEE Transactions on
  Information Forensics and Security}, vol.~18, pp. 4361--4376, 2023.

\bibitem{platerecognizer}
P.~Recognizer., ``Automatic license plate recognition - high accuracy alpr.''
  2022, https://platerecognizer.com/.

\bibitem{russakovsky2015imagenet}
O.~Russakovsky, J.~Deng, H.~Su, J.~Krause, S.~Satheesh, S.~Ma, Z.~Huang,
  A.~Karpathy, A.~Khosla, M.~Bernstein \emph{et~al.}, ``Imagenet large scale
  visual recognition challenge,'' \emph{International journal of computer
  vision}, vol. 115, pp. 211--252, 2015.

\bibitem{shafahi2019adversarial}
A.~Shafahi, M.~Najibi, M.~A. Ghiasi, Z.~Xu, J.~Dickerson, C.~Studer, L.~S.
  Davis, G.~Taylor, and T.~Goldstein, ``Adversarial training for free!''
  \emph{Advances in neural information processing systems}, vol.~32, 2019.

\bibitem{7958568}
R.~Shokri, M.~Stronati, C.~Song, and V.~Shmatikov, ``Membership inference
  attacks against machine learning models,'' in \emph{2017 IEEE Symposium on
  Security and Privacy (SP)}, 2017, pp. 3--18.

\bibitem{szegedy2013intriguing}
C.~Szegedy, W.~Zaremba, I.~Sutskever, J.~Bruna, D.~Erhan, I.~Goodfellow, and
  R.~Fergus, ``Intriguing properties of neural networks,'' \emph{arXiv preprint
  arXiv:1312.6199}, 2013.

\bibitem{tashiro2020diversity}
Y.~Tashiro, Y.~Song, and S.~Ermon, ``Diversity can be transferred: Output
  diversification for white-and black-box attacks,'' \emph{Advances in neural
  information processing systems}, vol.~33, pp. 4536--4548, 2020.

\bibitem{velickovic2017graph}
P.~Velickovic, G.~Cucurull, A.~Casanova, A.~Romero, P.~Lio, Y.~Bengio
  \emph{et~al.}, ``Graph attention networks,'' \emph{stat}, vol. 1050, no.~20,
  pp. 10--48\,550, 2017.

\bibitem{bounceattack}
J.~Wan, J.~Fu, L.~Wang, and Z.~Yang, ``Bounceattack: A query-efficient
  decision-based adversarial attack by bouncing into the wild,'' in \emph{2024
  IEEE Symposium on Security and Privacy (SP)}.\hskip 1em plus 0.5em minus
  0.4em\relax IEEE Computer Society, 2024, pp. 71--71.

\bibitem{wang2021deep}
M.~Wang and W.~Deng, ``Deep face recognition: A survey,''
  \emph{Neurocomputing}, vol. 429, pp. 215--244, 2021.

\bibitem{LineGraph}
E.~W. Weisstein, ``Line graph,'' 2024,
  \url{https://mathworld.wolfram.com/LineGraph.html}, From MathWorld--A Wolfram
  Web Resource.

\bibitem{5995566}
L.~Wolf, T.~Hassner, and I.~Maoz, ``Face recognition in unconstrained videos
  with matched background similarity,'' in \emph{CVPR 2011}, 2011, pp.
  529--534.

\bibitem{8429311}
S.~Yeom, I.~Giacomelli, M.~Fredrikson, and S.~Jha, ``Privacy risk in machine
  learning: Analyzing the connection to overfitting,'' in \emph{2018 IEEE 31st
  Computer Security Foundations Symposium (CSF)}, 2018, pp. 268--282.

\bibitem{yu2020cloudleak}
H.~Yu, K.~Yang, T.~Zhang, Y.-Y. Tsai, T.-Y. Ho, and Y.~Jin, ``Cloudleak:
  Large-scale deep learning models stealing through adversarial examples.'' in
  \emph{NDSS}, vol.~38, 2020, p. 102.

\bibitem{zhang2019graph}
S.~Zhang, H.~Tong, J.~Xu, and R.~Maciejewski, ``Graph convolutional networks: a
  comprehensive review,'' \emph{Computational Social Networks}, vol.~6, no.~1,
  pp. 1--23, 2019.

\bibitem{zhang2023no}
Z.~Zhang, C.~Gong, Y.~Cai, Y.~Yuan, B.~Liu, D.~Li, Y.~Guo, and X.~Chen, ``No
  privacy left outside: On the (in-) security of tee-shielded dnn partition for
  on-device ml,'' in \emph{2024 IEEE Symposium on Security and Privacy
  (SP)}.\hskip 1em plus 0.5em minus 0.4em\relax IEEE Computer Society, 2023,
  pp. 52--52.

\bibitem{zhang2020dynamic}
Z.~Zhang, Y.~Li, Y.~Guo, X.~Chen, and Y.~Liu, ``Dynamic slicing for deep neural
  networks,'' in \emph{Proceedings of the 28th ACM Joint Meeting on European
  Software Engineering Conference and Symposium on the Foundations of Software
  Engineering}, 2020, pp. 838--850.

\bibitem{zhang2023fedslice}
Z.~Zhang, Y.~Li, B.~Liu, Y.~Cai, D.~Li, Y.~Guo, and X.~Chen, ``Fedslice:
  Protecting federated learning models from malicious participants with model
  slicing,'' in \emph{2023 IEEE/ACM 45th International Conference on Software
  Engineering (ICSE)}.\hskip 1em plus 0.5em minus 0.4em\relax IEEE, 2023, pp.
  460--472.

\bibitem{zhang2022remos}
Z.~Zhang, Y.~Li, J.~Wang, B.~Liu, D.~Li, Y.~Guo, X.~Chen, and Y.~Liu, ``Remos:
  Reducing defect inheritance in transfer learning via relevant model
  slicing,'' in \emph{Proceedings of the 44th International Conference on
  Software Engineering}, 2022, pp. 1856--1868.

\end{thebibliography}

\newpage
\appendices
\section{Meta-Review}
\subsection{Summary}
The paper introduces a defense mechanism called Query Provenance Analysis (QPA) to address query-based black-box adversarial attacks on machine learning models. While traditional Stateful Defense Models (SDMs) are often vulnerable to adaptive attacks like Oracle-guided Adaptive Rejection Sampling (OARS), QPA improves defense by analyzing the relationships between sequential queries through a query provenance graph. This graph links each query to its most similar previous query, enabling the detection of adversarial patterns based on statistical features and query membership. Experimental results across multiple datasets demonstrate that QPA significantly outperforms existing methods in terms of defense effectiveness, detection latency, and throughput, offering robust protection against both non-adaptive and adaptive attacks.

\subsection{Scientific Contributions}
\begin{itemize}
    \item Creates a New Tool to Enable Future Science
    \item Provides a Valuable Step Forward in an Established Field
\end{itemize}

\subsection{Reasons for Acceptance}
\begin{enumerate}
    \item The paper proposes a new tool to detect black-box attacks and does an extensive evaluation of different aspects of the defense. The proposed tool has a higher efficiency and more effect than existing tools. 
    \item The novel approach of considering not only the current query but also the history of queries, combined with using a graph structure to represent these queries and employing GNN models to handle the generated graph for identifying attacking behavior, is both intriguing and effective.
\end{enumerate}

\end{document}